\documentclass[oupdraft]{sysbio}
\usepackage[T1]{fontenc}
\usepackage[dvipsnames,rgb]{xcolor}
\usepackage{times}

\usepackage[colorlinks=true, allcolors=blue]{hyperref}
\usepackage{cleveref}
\usepackage{nameref}

\usepackage[export]{adjustbox}
\usepackage{graphicx}
\usepackage[caption=false]{subfig}
\usepackage{array}
\usepackage{booktabs}

\usepackage{url}
\usepackage{tikz}
\usetikzlibrary{decorations.pathmorphing, decorations.pathreplacing, positioning,shapes.geometric}
\tikzset{every picture/.style={/utils/exec={\sffamily}}}

\usepackage{algorithm}
\usepackage{algpseudocode}

\usepackage{dsfont}

\renewcommand{\v}{$\boldsymbol{v}$}
\DeclareRobustCommand{\sbseries}{\fontseries{sb}\selectfont}
\DeclareTextFontCommand{\textsb}{\sbseries}

\begin{document}

\title{Phylo2Vec: a vector representation for binary trees}

\author{Matthew J Penn$^{1\ast}$, Neil Scheidwasser$^{2\ast}$, Mark P Khurana$^{2}$, David A Duch\^ene$^{1}$, Christl A Donnelly$^{1, 3}$, and Samir Bhatt$^{2, 4}$\\[4pt]
\textit{$^{1}$Department of Statistics, University of Oxford, Oxford, United Kingdom}
\\
\textit{$^{2}$Section of Epidemiology, University of Copenhagen, Copenhagen, Denmark}
\\
\textit{$^{3}$Pandemic Sciences Institute, University of Oxford, Oxford, United Kingdom}
\\
\textit{$^{4}$MRC Centre for Global Infectious Disease Analysis, Imperial College London, London, United Kingdom}
\\
\textit{$^{\ast}$Equal contribution}
\\[2pt]
Correspondence: \textit{neil.clow@sund.ku.dk}}

\markboth%
{Penn et al.}
{Phylo2Vec}

\maketitle

\begin{abstract}
{Binary phylogenetic trees inferred from biological data are central to understanding the shared history among evolutionary units. However, inferring the placement of latent nodes in a tree is computationally expensive. State-of-the-art methods rely on carefully designed heuristics for tree search, using different data structures for easy manipulation (e.g., classes in object-oriented programming languages) and readable representation of trees (e.g., Newick-format strings). Here, we present Phylo2Vec, a parsimonious encoding for phylogenetic trees that serves as a unified approach for both manipulating and representing phylogenetic trees. Phylo2Vec maps any binary tree with $n$ leaves to a unique integer vector of length $n-1$. The advantages of Phylo2Vec are fourfold: i) fast tree sampling, (ii) compressed tree representation compared to a Newick string, iii) quick and unambiguous verification if two binary trees are identical topologically, and iv) systematic ability to traverse tree space in very large or small jumps. As a proof of concept, we use Phylo2Vec for maximum likelihood inference on five real-world datasets and show that a simple hill-climbing-based optimisation scheme can efficiently traverse the vastness of tree space from a random to an optimal tree.}
{phylogenetics, representation, binary trees, optimisation}
\end{abstract}
\newline

Phylogenetic trees are a fundamental tool for depicting evolutionary processes, whether linguistic (evolution of different languages and language families) or biological (evolution of biological entities). Within the biological sciences, phylogenetic trees are integral to multiple research domains, including evolution~\citep{morlon_potts}, conservation~\citep{rolland_cadotte}, and epidemiology, where they allow us to better understand infectious disease transmission dynamics~\citep{ypma_ballegooijen, faria2021}.

A multitude of computer-readable formats have been proposed to store and represent (binary) phylogenetic trees. Although basic data structures such as arrays or linked lists can be used for this purpose, the Newick format, as outlined by~\citet{olsen1990} and~\citet{felsenstein2004}, has emerged as the standard notation. This format characterises a tree through a string of nested parentheses. Each parenthesis encloses a pair of leaf nodes or subtrees, separated by a comma. Additional metadata such as branch lengths can be added after a colon which follows the leaf node or subtree. Although a compact and intuitive notation, several limitations exist. First, comparing (large) trees using the Newick format can be difficult for human readers, especially as isomorphic trees can be obtained by permuting nodes or subtrees within a set of parentheses. Second, verifying that two trees are identical from Newick strings requires additional steps, as two identical trees need not have the identical Newick string. Alternative, bijective representations do exist. For example, several methods have been proposed to assign unique integers to binary trees with unlabelled~\citep{rotem1978, proskurowski1980}, fully labelled~\citep{knuth1973}, and partially labelled nodes~\citep{rohlf1983} (only leaf nodes). More recently, \citet{palacios2022} investigated several enumeration strategies of binary trees compatible with a perfect phylogeny. However, as mentioned by~\citet{rohlf1983}, using single-integer representations for downstream phylogenetic analyses is computationally difficult for large trees due to the factorial rate at which the size of binary tree space grows~\citep{Cavalli-Sforza1967-xh}. Conversely, several vector representations such as graph polynomials~\citep{liu2021, liu2022} and the compact bijective ladderized vector~\citep{voznica2022} were introduced as a support for model selection and estimation of evolutionary or epidemiological parameters. Other vector representations of tree topology, such as pair matchings~\citep{diaconis1998} and F matrices~\citep{kim2020}, focus on the polynomial-time computation of the distance between any two trees (to measure similarity). However, methods for systematically sampling random trees or changing tree topology with respect to an objective function by leveraging such vector representations have been understudied. In particular, creating sampling schemes (as done in Bayesian frameworks such as \texttt{BEAST}~\citep{Drummond2007-za, bouckaert2014} and \texttt{MrBayes}~\citep{huelsenbeck2001}) around standard tree arrangements is non-trivial, and, although inferring phylogenetic trees is a common task in evolutionary biology, tree search using any optimality criteria (including maximum likelihood) is NP-hard~\citep{Roch2006-qs}. Another critical challenge is the size of the tree space: for a tree with \textit{n} leaves, there are $(2n-3)\cdot(2n-5)\cdot\ldots\cdot5\cdot3\cdot1$ possible rooted binary trees~\citep{Cavalli-Sforza1967-xh}. Lastly, optimisation-based approaches often face a jagged ``loss" landscape containing many trees with the same criterion score~\citep{sanderson2011}. When considering inference, the choice of representation can be particularly relevant for application to real phylogenetic problems. For example, an application of the approach we introduce here can be used for continuous relaxation and gradient descent under the minimum evolution criterion~\citep{penn2023}. For large phylogenies, the use of an efficient representation such as the compact bijective ladderized vector~\citep{voznica2022} has proven effective for deep learning-based, likelihood-free, inference~\citep{thompson2024} or diversification inference~\citep{lambert2023}.

To overcome these limitations, we introduce Phylo2Vec, a new representation for any binary tree. In this framework, the topology of a binary tree can be completely described by a single integer vector \v~of dimension $n-1$, where $n$ is the number of leaves in the tree. The vector's construction is intrinsically related to the branching pattern of the tree, and is defined by a simple constraint: $v_j \in \{0, 1, \ldots, 2(j-1)\}$ for all $ j \in \{1,\ldots,n-1\}$. The approach we present here is most similar to that previously introduced by~\citet{rohlf1983}, but we focus on the integer representation and its mathematical properties, rather than counting or labelling trees. Additionally, this formulation naturally offers a new measure of distance between trees (e.g., by comparing two vectors using the Hamming distance) and yields a new mechanism to explore tree space which diverges from traditional heuristics such as subtree, prune and regraft (SPR). To further demonstrate its utility, as a proof of concept, we apply Phylo2Vec to several phylogenetic inference problems, where the task is to find an optimal tree given a set of genetic sequences using maximum likelihood estimation. While state-of-the-art frameworks for phylogenetic inference typically rely on search heuristics based on deterministic tree arrangements, Phylo2Vec provides the first steps to a more systematic criterion for optimisation.

\section{Materials and Methods}
The goal of this project was to develop a bijection (i.e., a one-to-one correspondence) between the set of binary rooted trees with $n$ leaves to a constrained set of integer vectors of length $n-1$. We first describe an intuitive but incomplete (as not bijective) integer representation of trees as birth processes. Second, we define and characterise Phylo2Vec as a bijective generalisation of this first representation and formalise its properties. Third, we showcase the utility of Phylo2Vec by applying the representation for MLE-based phylogenetic inference on empirical datasets. 

Our construction draws from an existing method of assigning integer counts to trees~\citep{rohlf1983}, although we focus on vector representations. It is distinct from~\citet{rohlf1983} in labelling the tree edges, motivated by a simple and intuitive representation of birth processes. By applying this encoding to rooted binary trees, we are able to move around tree space similarly to subtree-prune and regraft methods.
Furthermore, we provide a rigorous proof of its bijectivity alongside a range of algorithms (all implemented into a Python package) which allows researchers to build on the phylogenetic optimisation algorithm we present here. Thus, we provide a significantly different method from those proposed previously~\citep{rohlf1983}, by focusing our efforts toward practical transitions in tree space.

\subsection{An incomplete integer representation of trees as birth processes}\label{sec:ordered}
Let $\mathcal{T}$ denote a rooted phylogenetic tree with $n$ leaf nodes representing (biological) taxa, and $\mathcal{D}$ symbolise a key-value mapping (or dictionary) which associates a nonnegative integer (the keys) to each leaf node (the values).

Using this mapping, \emph{for a subset of all trees}, we can summarise their topology using an integer vector \v~of size $n-1$ such that:

\begin{equation}
    v_j \in \{0, 1, \ldots, j-1\} \quad \forall j \in \{1, \ldots, n-1\}
    \label{eq:ordered}
\end{equation}

The construction of this vector is inspired by birth processes: assuming a two-leaf tree with leaves labelled 0 and 1, we process \v~from left to right. For each $j \in \{1, \ldots, n-1\}$, $v_j$ (hereinafter noted as \v$[j]$) denotes the addition of leaf $j$ such that, at iteration $j$, leaf $j$ forms a cherry with leaf \v$[j]$. In other words, the branch leading to leaf \v$[j]$ ``gives birth" to leaf $j$. Figure~\ref{fig:ordered} illustrates algorithms to convert a tree to a vector and \textit{vice versa}.

Although a simple representation of tree topology, it is easy to see from Equation~\ref{eq:ordered} that this construction is incomplete. Indeed, there are $j$ possible values for any \v$[j]$, and thus, for $n$ leaves, there are $1 \cdot 2 \cdot \ldots \cdot (n-1) = (n-1)!$ possible vectors, which is less than the number of binary rooted trees, $(2n-3)!!$ (where $!!$ denotes the semifactorial)~\citep{Cavalli-Sforza1967-xh,Felsenstein1978-qc,diaconis1998}. This discrepancy stems from the assumptions of this construction, whereby a new leaf $j$ has to form a cherry with a previously processed leaf $0, 1, \ldots, j-1$. For instance, leaf 2 has to form a cherry with either leaf 0 or 1, but cannot be an outgroup of the (0, 1) subtree. We thus denote trees that follow this incomplete construction of tree space as ``ordered" trees, as they require a precise ordering of the leaf nodes.

\begin{figure}[htbp]
\centering
\fontsize{9}{10.8}\selectfont
\sffamily
\setlength{\fboxsep}{0pt}
\setlength{\tabcolsep}{3pt}

\subfloat[]{
\adjustbox{valign=b}{
\fbox{
\begin{tabular}{l >{\centering\arraybackslash}m{3.55cm} m{4.1cm}}
     \textsb{Input} & Newick = (((0,3)4,2)5,1)6; \\
           & \begin{tikzpicture}[thick, scale=0.75, every node/.style={scale=0.9}]
        \node[circle, fill=lime] (6) at (1.5, 2) {};
        \node[circle, fill=lime] (5) at (0.5, 1) {};
        \node[circle, fill=lime] (4) at (1, 1.5) {};
        \node[circle, fill=OliveGreen] (0) at (0, 0) {0};
        \node[circle, fill=OliveGreen] (2) at (2, 0) {2};
        \node[circle, fill=OliveGreen] (3) at (1, 0) {3};
        \node[circle, fill=OliveGreen] (1) at (3, 0) {1};

        \draw (6) -- (4);
        \draw (5) -- (4);
        \draw (6) -- (1);
        \draw (4) -- (2);
        \draw (5) -- (0);
        \draw (5) -- (3);
    \end{tikzpicture} \\
     Step 1 & Next leaf: 1 & v = [0] \\
           & (((\textsb{0},\textcolor{gray}{3),2),}\textsb{1}); \\ 
           & \begin{tikzpicture}[thick, scale=0.75, every node/.style={scale=0.9}]
        \node[circle, fill=lime] (2) at (1.5, 2) {};
        \node[circle, fill=OliveGreen] (0) at (0, 0) {0};
        \node[circle, fill=red] (1) at (3, 0) {1};
        \draw[red] (2) -- node[above right] {{}} (1);
        \draw (2) -- node[above left] {{}} (0);
    \end{tikzpicture}                          & There is only 1 possible arrangement of a 2-leaf tree. \newline $\rightarrow$ v[1] is always 0. \\
           &  &  \\
     Step 2 & Next leaf: 2 & v = [0, 0] \\
            & (((\textsb{0},\textcolor{gray}{3)}\textsb{2})1); \\
            & \begin{tikzpicture}[thick, scale=1, every node/.style={scale=0.9}]
        \node[circle, fill=lime] (6) at (1.5, 2) {};
        \node[circle, fill=lime] (5) at (0.5, 1) {};
        \node[circle, fill=OliveGreen] (0) at (0, 0) {0};
        \node[circle, fill=red] (2) at (1, 0) {2};
        \node[circle, fill=OliveGreen] (1) at (3, 0) {1};

        \draw (6) -- (1);
        \draw (6) -- node[above right] {{}} (5);
        \draw (5) -- node[above left] {{}} (0);
        \draw[red] (5) -- node[above right] {{}} (2);
    \end{tikzpicture}             & Leaves 2 and 0 form a cherry \newline in a 3-leaf tree \newline $\rightarrow$ v[2] = 0 \\
     Step 3 & Next leaf: 3 & v = [0, 0, 0] \\
            & (((\textsb{0},\textsb{3})2)1); \\
            & \begin{tikzpicture}[thick, scale=1, every node/.style={scale=0.9}]
        \node[circle, fill=lime] (6) at (1.5, 2) {};
        \node[circle, fill=lime] (5) at (0.5, 1) {};
        \node[circle, fill=lime] (4) at (1, 1.5) {};
        \node[circle, fill=OliveGreen] (0) at (0, 0) {0};
        \node[circle, fill=OliveGreen] (2) at (2, 0) {2};
        \node[circle, fill=red] (3) at (1, 0) {3};
        \node[circle, fill=OliveGreen] (1) at (3, 0) {1};

        \draw (6) -- (4);
        \draw (5) -- (4);
        \draw (6) -- (1);
        \draw (4) -- (2);
        \draw (5) -- (0);
        \draw[red] (5) -- (3);
    \end{tikzpicture}             & Leaves 3 and 0 form a cherry \newline in a 4-leaf tree \newline $\rightarrow$ v[3] = 0 \\
\end{tabular}
}
}
}
\subfloat[]{
\adjustbox{valign=b}{
\fbox{
    \begin{tabular}{l l c}
         \textsb{Input} & \v~=~[0, 0, 1] \\
         Step 1 & \v[1] = [0] & Initial tree: \\
                &             & 2 leaves: 0, 1 \\
                &             & 1 internal node \\
         & & \begin{tikzpicture}[thick, scale=1, every node/.style={scale=1}]
            \node[circle, fill=lime] (2) at (1, 1) {};
            \node[circle, fill=OliveGreen] (0) at (0, 0) {0};
            \node[circle, fill=OliveGreen] (1) at (2, 0) {1};
            
            \draw (2) -- node[above left] {{0}} (0);
            \draw (2) -- node[above right] {{1}} (1);
        \end{tikzpicture} \\
        Step 2 & \v[2] = 0 & Split branch 0, yield leaf 2  \\
        & &
        \begin{tikzpicture}[thick, scale=1, every node/.style={scale=1}]
            \node[circle, fill=lime] (3) at (0.5, 1) {};
            \node[circle, fill=lime] (4) at (1, 2) {};
            \node[circle, fill=OliveGreen] (0) at (0, 0) {0};
            \node[circle, fill=OliveGreen] (1) at (2, 0) {1};
            \node[circle, fill=OliveGreen] (2) at (1, 0) {2};
            
            \draw (4) -- node[above right] {{1}} (1);
            \draw (4) -- (3);
            \draw (3) -- node[above left] {{0}} (0);
            \draw (3) -- node[above right] {{2}} (2);
        \end{tikzpicture}
        \\
        Step 3 & \v[3] = 1 & Split branch 1, yield leaf 3 \\
        & & \begin{tikzpicture}[thick, scale=1, every node/.style={scale=1}]
            \node[circle, fill=lime] (6) at (1.5, 2) {};
            \node[circle, fill=lime] (5) at (0.5, 1) {};
            \node[circle, fill=lime] (4) at (2.5, 1) {};
            \node[circle, fill=OliveGreen] (0) at (0, 0) {0};
            \node[circle, fill=OliveGreen] (2) at (1, 0) {2};
            \node[circle, fill=OliveGreen] (3) at (2, 0) {3};
            \node[circle, fill=OliveGreen] (1) at (3, 0) {1};

            \draw (6) -- (4);
            \draw (6) -- (5);
            \draw (4) -- node[above right] {{1}} (1);
            \draw (4) -- node[above left] {{3}} (3);
            \draw (5) -- node[above left] {{0}} (0);
            \draw (5) -- node[above right] {{2}} (2);
        \end{tikzpicture} \\
        Final & \v~= [0, 0, 1] & Name the ancestors (LIFO) \\
        & & Newick: ((1,3)4,(0,2)5)6; \\
        & & \begin{tikzpicture}[thick, scale=1, every node/.style={scale=0.9}]
            \node[circle, fill=lime] (6) at (1.5, 2) {6};
            \node[circle, fill=lime] (5) at (0.5, 1) {5};
            \node[circle, fill=lime] (4) at (2.5, 1) {4};
            \node[circle, fill=OliveGreen] (0) at (0, 0) {0};
            \node[circle, fill=OliveGreen] (2) at (1, 0) {2};
            \node[circle, fill=OliveGreen] (3) at (2, 0) {3};
            \node[circle, fill=OliveGreen] (1) at (3, 0) {1};

            \draw (6) -- (4);
            \draw (6) -- (5);
            \draw (4) -- node[above right] {{1}} (1);
            \draw (4) -- node[above left] {{3}} (3);
            \draw (5) -- node[above left] {{0}} (0);
            \draw (5) -- node[above right] {{2}} (2);
        \end{tikzpicture} \\
    \end{tabular}
}
}
}
\rmfamily
\setlength{\fboxsep}{3pt}
\setlength{\tabcolsep}{6pt}
\caption{An incomplete integer representation of tree topology as birth processes. \textsb{(a)} Labelling a tree as an ordered vector: example for \v~$=[0, 0, 0]$. We process leaves in ascending order. For each leaf $j$, we retrieve its sibling (or adjacent tip) in the Newick string, ignoring leaves > $j$. The adjacent tip corresponds to \v$[j]$. \textsb{(b)} Recovering a tree from an ordered vector: example for \v~$= [0, 0, 1]$. We process \v~from left to right. Ancestors are named in last-in-first-out (LIFO) fashion: The ancestor of the last added leaf $L-1$ (here, leaf 3) is named $L$ (here, 4), the ancestor of the second-to-last added leaf $L-2$ (here, leaf 2) is named $L+1$ (here, 5) etc. In both cases, the lengths of the edges are arbitrary.}
\label{fig:ordered}
\end{figure}

\subsection{Phylo2Vec}
In this section, we define and formalise the properties of Phylo2Vec, an integer vector representation that extends the formulation presented above to be valid for any rooted binary tree. To ensure bijectivity to this space, we need the vector \v~to satisfy the following constraints:


\begin{equation}
    v_j \in \{0, 1, \ldots, 2(j-1)\} ~\forall j \in \{1,\ldots,n-1\}
    \label{eq:phylo2vec}
\end{equation}

We say \v~$\in \mathbb{V}$ if Equation~\ref{eq:phylo2vec} is satisfied. For this representation, there are $2j-1$ entries for any position $j$. Therefore, the number of possible vectors matches the number of possible binary rooted trees:

\begin{equation*}
    \prod_{j=1}^{n-1} (2j-1) = (2(n-1)-1)!! = (2n-3)!!
\end{equation*}

From this observation, we can prove the bijectivity of the mapping simply by showing injectivity - that is, that no two distinct vectors \v~and $\boldsymbol{w}$ lead to the same tree. A proof is presented in the Appendix (\nameref{sec:phylo2vec}). Briefly, our proof relies on the fact that certain properties of pairs of nodes are preserved throughout the construction process - namely, that the most recent common ancestor (MRCA) of a pair of nodes is unchanged (once both nodes have been added to the tree) and that if one node is the ancestor of another at some stage of the construction process, then this remains true in the final tree. Then, if $\mathcal{T}$ and $\mathcal{T}'$ are the trees resulting from different vectors \v~and \v', respectively, we choose the smallest $i$ such that $v_i \neq v'_i$. By considering the sets of leaf nodes descended from the edge to which $i$ is added, we can show that the addition of node $i$ causes a pair of nodes to either have a different MRCA or a different ancestral relationship. Therefore, since these properties are preserved throughout the construction process, we must have $\mathcal{T} \neq \mathcal{T}'$. This shows the injectivity of our mapping, with bijectivity following from the fact that the number of trees is the same as the number of possible vectors \v. 

\subsubsection{Recovering a tree from a Phylo2Vec vector}
Building a binary tree from \v~follows closely the algorithm in Figure~\ref{fig:ordered}, but incorporates two additional requirements. First, we start from a two-leaf tree, whereby the leaves are labelled 0 and 1. The branches that lead to leaves 0, 1 are also labelled 0, 1, respectively. Second, we draw an additional node (called the ``extra root") which is initially connected to the root by a branch labelled 2 (see second row in Fig.~\ref{fig:to_newick}).

\begin{figure}[htbp]
    \centering
    \setlength{\tabcolsep}{2pt}
    \sffamily
    \small
\subfloat[]{
\adjustbox{valign=t}{
    \begin{tabular}{p{3cm}@{\hspace{-1cm}}c}
        \toprule
         \textsb{Recovering a tree} \newline Ex: \v~=~[0, 2, 2, 5, 2] \\
         \midrule
         \textsb{Step 1} \newline \v[1] = [0] \newline Initial tree: \newline 2 leaves: 0, 1 \newline 1 internal node \newline 1 extra root (R) & \adjustbox{valign=t}{\begin{tikzpicture}[thick, scale=0.85, every node/.style={scale=0.85}]
            \node[circle, fill=lightgray] (R) at (1, 2) {R};
            \node[circle, fill=lime] (2) at (1, 1) {};
            \node[circle, fill=OliveGreen] (0) at (0, 0) {0};
            \node[circle, fill=OliveGreen] (1) at (2, 0) {1};
            
            \draw (R) -- node[right] {{2}} (2);
            \draw (2) -- node[left] {{0}} (0);
            \draw (2) -- node[right] {{1}} (1);
        \end{tikzpicture}} \\
        \midrule
        \textsb{Step 2} \newline \v[2] = 2 \newline Split branch 2 \newline yield leaf 2 & \adjustbox{valign=t}{\begin{tikzpicture}[thick, scale=0.85, every node/.style={scale=0.85}]
            \node[circle, fill=lightgray] (R) at (1, 2) {R};
            \node[circle, fill=lime] (4) at (1, 1) {};
            \node[circle, fill=lime] (3) at (0.5, 0.5) {};
            \node[circle, fill=OliveGreen] (0) at (0, 0) {0};
            \node[circle, fill=OliveGreen] (1) at (1, 0) {1};
            \node[circle, fill=OliveGreen] (2) at (2, 0) {2};
            
            \draw (R) -- node[right] {{4}} (4);
            \draw (4) -- node[above right] {{2}} (2);
            \draw (4) -- node[above left] {{3}} (3);
            \draw (3) -- node[above left] {{0}} (0);
            \draw (3) -- node[above right] {{1}} (1);
        \end{tikzpicture}}\\
        \midrule
        \textsb{Step 3} \newline \v[3] = 2 \newline Split branch 2 \newline  yield leaf 3 & \adjustbox{valign=t}{\begin{tikzpicture}[thick, scale=0.85, every node/.style={scale=0.85}]
            \node[circle, fill=OliveGreen] (0) at (0, 0) {0};
            \node[circle, fill=OliveGreen] (1) at (1, 0) {1};
            \node[circle, fill=OliveGreen] (2) at (3, 0) {2};
            \node[circle, fill=OliveGreen] (3) at (2, 0) {3};
            \node[circle, fill=lime] (5) at (0.5, 1) {};
            \node[circle, fill=lime] (4) at (2.5, 1) {};
            \node[circle, fill=lime] (6) at (1.5, 1.75) {};
            \node[circle, fill=lightgray] (R) at (1.5, 2.5) {R};
            
            \draw (R) -- node[right] {{6}} (6);
            \draw (6) -- node[above right] {{4}} (4);
            \draw (6) -- node[above left] {{5}} (5);
            \draw (5) -- node[left] {{0}} (0);
            \draw (5) -- node[right] {{1}} (1);
            \draw (4) -- node[right] {{2}} (2);
            \draw (4) -- node[left] {{3}} (3);
        \end{tikzpicture}} \\
        \midrule
        \textsb{Step 4} \newline \v[4] = 5 \newline Split branch 5 \newline  yield leaf 4 & \adjustbox{valign=t}{\begin{tikzpicture}[thick, scale=0.85, every node/.style={scale=0.85}]
            \node[circle, fill=OliveGreen] (0) at (0, 0) {0};
            \node[circle, fill=OliveGreen] (1) at (1, 0) {1};
            \node[circle, fill=OliveGreen] (2) at (4, 0) {2};
            \node[circle, fill=OliveGreen] (3) at (3, 0) {3};
            \node[circle, fill=OliveGreen] (4) at (2, 0) {4};
            \node[circle, fill=lime] (5) at (3.5, 1) {};
            \node[circle, fill=lime] (7) at (1.25, 1.375) {};
            \node[circle, fill=lime] (6) at (0.5, 1) {};
            \node[circle, fill=lime] (8) at (2, 1.75) {};
            \node[circle, fill=lightgray] (R) at (2, 2.5) {R};
            
            \draw (R) -- node[right] {{8}} (8);
            \draw (8) -- node[above left] {{7}} (7);
            \draw (8) -- node[above right] {{5}} (5);
            \draw (7) -- node[above left] {{6}} (6);
            \draw (7) -- node[right] {{4}} (4);
            \draw (6) -- node[left] {{0}} (0);
            \draw (6) -- node[right] {{1}} (1);
            \draw (5) -- node[left] {{3}} (3);
            \draw (5) -- node[right] {{2}} (2);
        \end{tikzpicture}} \\
        \midrule
        \textsb{Step 5} \newline \v[5] = 2 \newline Split branch 2 \newline  yield leaf 5 & \adjustbox{valign=t}{\begin{tikzpicture}[thick, scale=0.85, every node/.style={scale=0.85}]
            \node[circle, fill=OliveGreen] (0) at (0, 0) {0};
            \node[circle, fill=OliveGreen] (1) at (1, 0) {1};
            \node[circle, fill=OliveGreen] (2) at (5, 0) {2};
            \node[circle, fill=OliveGreen] (5) at (4, 0) {5};
            \node[circle, fill=OliveGreen] (4) at (2, 0) {4};
            \node[circle, fill=OliveGreen] (3) at (3, 0) {3};
            \node[circle, fill=lime] (6) at (4.5, 1) {};
            \node[circle, fill=lime] (7) at (3.5, 1.5) {};
            \node[circle, fill=lime] (8) at (0.5, 1) {};
            \node[circle, fill=lime] (9) at (1.5, 1.5) {};
            \node[circle, fill=lime] (10) at (2.5, 2) {};
            \node[circle, fill=lightgray] (R) at (2.5, 2.75) {R};
            
            \draw (R) -- node[right] {{10}} (10);
            \draw (10) -- node[above left] {{9}} (9);
            \draw (10) -- node[above right] {{7}} (7);
            \draw (9) -- node[above left] {{8}} (8);
            \draw (9) -- node[right] {{4}} (4);
            \draw (8) -- node[left] {{0}} (0);
            \draw (8) -- node[right] {{1}} (1);
            \draw (7) -- node[left] {{3}} (3);
            \draw (7) -- node[above right] {{6}} (6);
            \draw (6) -- node[right] {{2}} (2);
            \draw (6) -- node[left] {{5}} (5);
        \end{tikzpicture}} \\
        \midrule
        \textsb{Final} \newline Remove the \newline extra root. \newline Name the \newline ancestors. & \adjustbox{valign=t}{\begin{tikzpicture}[thick, scale=0.85, every node/.style={scale=0.85}]
            \node[circle, fill=OliveGreen] (0) at (0, 0) {0};
            \node[circle, fill=OliveGreen] (1) at (1, 0) {1};
            \node[circle, fill=OliveGreen] (2) at (5, 0) {2};
            \node[circle, fill=OliveGreen] (5) at (4, 0) {5};
            \node[circle, fill=OliveGreen] (4) at (2, 0) {4};
            \node[circle, fill=OliveGreen] (3) at (3, 0) {3};
            \node[circle, fill=lime] (6) at (4.5, 1) {6};
            \node[circle, fill=lime] (7) at (3.5, 1.5) {7};
            \node[circle, fill=lime] (8) at (0.5, 1) {8};
            \node[circle, fill=lime] (9) at (1.5, 1.5) {9};
            \node[circle, fill=lime] (10) at (2.5, 2) {10};
            
            \draw (10) -- node[above left] {{9}} (9);
            \draw (10) -- node[above right] {{7}} (7);
            \draw (9) -- node[above left] {{8}} (8);
            \draw (9) -- node[right] {{4}} (4);
            \draw (8) -- node[left] {{0}} (0);
            \draw (8) -- node[right] {{1}} (1);
            \draw (7) -- node[left] {{3}} (3);
            \draw (7) -- node[above right] {{6}} (6);
            \draw (6) -- node[right] {{2}} (2);
            \draw (6) -- node[left] {{5}} (5);
        \end{tikzpicture}} \\
        \bottomrule
    \end{tabular}
}
}
\subfloat[]{
\adjustbox{valign=t}{
    \begin{tabular}{p{4cm}c}
        \toprule
        \textsb{Labelling branches} \newline Ex:  \v~=~[0, 2, 2, 5, 2]
        \\
        \midrule
        \textsb{Step 1: \newline Leaf branches \newline \& extra root} & \adjustbox{valign=t}{\begin{tikzpicture}[thick, scale=0.9, every node/.style={scale=0.9}]
            \node[circle, fill=OliveGreen] (0) at (0, 0) {0};
            \node[circle, fill=OliveGreen] (1) at (1, 0) {1};
            \node[circle, fill=OliveGreen] (2) at (5, 0) {2};
            \node[circle, fill=OliveGreen] (5) at (4, 0) {5};
            \node[circle, fill=OliveGreen] (4) at (2, 0) {4};
            \node[circle, fill=OliveGreen] (3) at (3, 0) {3};
            \node[circle, fill=lime] (6) at (4.5, 1) {};
            \node[circle, fill=lime] (7) at (3.5, 1.5) {};
            \node[circle, fill=lime] (8) at (0.5, 1) {};
            \node[circle, fill=lime] (9) at (1.5, 1.5) {};
            \node[circle, fill=lime] (10) at (2.5, 2) {};
            \node[circle, fill=lightgray] (R) at (2.5, 2.75) {R};
            
            \draw (R) -- node[right] {{10}} (10);
            \draw (10) -- node[above left] {{}} (9);
            \draw (10) -- node[above right] {{}} (7);
            \draw (9) -- node[above left] {{}} (8);
            \draw (9) -- node[right] {{{\fontsize{13}{15.6}\selectfont\textsb{4}}}} (4);
            \draw (8) -- node[left] {{{\fontsize{13}{15.6}\selectfont\textsb{0}}}} (0);
            \draw (8) -- node[right] {{{\fontsize{13}{15.6}\selectfont\textsb{1}}}} (1);
            \draw (7) -- node[left] {{{\fontsize{13}{15.6}\selectfont\textsb{3}}}} (3);
            \draw (7) -- node[above right] {{}} (6);
            \draw (6) -- node[right] {{{\fontsize{13}{15.6}\selectfont\textsb{2}}}} (2);
            \draw (6) -- node[left] {{{\fontsize{13}{15.6}\selectfont\textsb{5}}}} (5);
        \end{tikzpicture}} \\
        \midrule
        \textsb{Step 2: \newline Internal branches} \newline Two cherries. \newline Find the cherry with \newline the highest label (here: \textsb{5}). \newline Label its parent branch as \textsb{6}. \newline Prune leaf \textsb{5}.  & \adjustbox{valign=t}{\begin{tikzpicture}[thick, scale=0.9, every node/.style={scale=0.9}]
            \node[circle, fill=OliveGreen] (0) at (0, 0) {0};
            \node[circle, fill=OliveGreen] (1) at (1, 0) {1};
            \node[circle, fill=OliveGreen] (2) at (5, 0) {2};
            \node[circle, draw=OliveGreen] (5) at (4, 0) {5};
            \node[circle, fill=OliveGreen] (4) at (2, 0) {4};
            \node[circle, fill=OliveGreen] (3) at (3, 0) {3};
            \node[rectangle, fill=lime, scale=2] (6) at (4.5, 1) {};
            \node[circle, fill=lime] (7) at (3.5, 1.5) {};
            \node[rectangle, fill=lime, scale=2] (8) at (0.5, 1) {};
            \node[circle, fill=lime] (9) at (1.5, 1.5) {};
            \node[circle, fill=lime] (10) at (2.5, 2) {};
            \node[circle, fill=lightgray] (R) at (2.5, 2.75) {};
            
            \draw (R) -- node[right] {{10}} (10);
            \draw (10) -- node[above left] {{}} (9);
            \draw (10) -- node[above right] {{}} (7);
            \draw (9) -- node[above left] {{}} (8);
            \draw (9) -- node[right] {{4}} (4);
            \draw (8) -- node[left] {{0}} (0);
            \draw (8) -- node[right] {{1}} (1);
            \draw (7) -- node[left] {{3}} (3);
            \draw (7) -- node[above] {{{\fontsize{14}{16.8}\selectfont\textsb{6}}}} (6);
            \draw (6) -- node[right] {{2}} (2);
            \draw[dotted] (6) -- node[left] {{5}} (5);
        \end{tikzpicture}} \\
        \midrule
        Two cherries. \newline Find the cherry with \newline the highest label (here: \textsb{3}). \newline Label its parent branch as \textsb{7}. \newline Prune leaf \textsb{3}. & \adjustbox{valign=t}{\begin{tikzpicture}[thick, scale=0.9, every node/.style={scale=0.9}]
            \node[circle, fill=OliveGreen] (0) at (0, 0) {0};
            \node[circle, fill=OliveGreen] (1) at (1, 0) {1};
            \node[circle, fill=OliveGreen] (2) at (5, 0) {2};
            \node[circle, fill=OliveGreen] (4) at (2, 0) {4};
            \node[circle, draw=OliveGreen] (3) at (3, 0) {3};
            \node[rectangle, fill=lime, scale=2] (7) at (3.5, 1.5) {};
            \node[rectangle, fill=lime, scale=2] (8) at (0.5, 1) {};
            \node[circle, fill=lime] (9) at (1.5, 1.5) {};
            \node[circle, fill=lime] (10) at (2.5, 2) {};
            \node[circle, fill=lightgray] (R) at (2.5, 2.75) {};
            
            \draw (R) -- node[right] {{10}} (10);
            \draw (10) -- node[above left] {{}} (9);
            \draw (10) -- node[below] {{{\fontsize{14}{16.8}\selectfont\textsb{7}}}} (7);
            \draw (9) -- node[above left] {{}} (8);
            \draw (9) -- node[right] {{4}} (4);
            \draw (8) -- node[left] {{0}} (0);
            \draw (8) -- node[right] {{1}} (1);
            \draw[dotted] (7) -- node[left] {{3}} (3);
            \draw (7) -- node[above right] {{2}} (2);
        \end{tikzpicture}} \\
        \midrule
        One cherry. \newline Label its parent branch as \textsb{8}. \newline Prune the cherry leaf with \newline the highest label = \textsb{1}. & \adjustbox{valign=t}{\begin{tikzpicture}[thick, scale=0.9, every node/.style={scale=0.9}]
            \node[circle, fill=OliveGreen] (0) at (0, 0) {0};
            \node[circle, draw=OliveGreen] (1) at (1, 0) {1};
            \node[circle, fill=OliveGreen] (2) at (5, 0) {2};
            \node[circle, fill=OliveGreen] (4) at (2, 0) {4};
            \node[rectangle, fill=lime, scale=2] (8) at (0.5, 1) {};
            \node[circle, fill=lime] (9) at (1.5, 1.5) {};
            \node[circle, fill=lime] (10) at (2.5, 2) {};
            \node[circle, fill=lightgray] (R) at (2.5, 2.75) {};
            
            \draw (R) -- node[right] {{10}} (10);
            \draw (10) -- node[above left] {{}} (9);
            \draw (9) -- node[above] {{{\fontsize{14}{16.8}\selectfont\textsb{8}}}} (8);
            \draw (9) -- node[right] {{4}} (4);
            \draw (8) -- node[left] {{0}} (0);
            \draw[dotted] (8) -- node[right] {{1}} (1);
            \draw (10) -- node[above right] {{2}} (2);
        \end{tikzpicture}} \\
        \midrule
        One cherry. \newline Label its parent branch as \textsb{9}. \newline Prune the cherry leaf with \newline the highest label = \textsb{4}. & \adjustbox{valign=t}{\begin{tikzpicture}[thick, scale=0.9, every node/.style={scale=0.9}]
            \node[circle, fill=OliveGreen] (0) at (0, 0) {0};
            \node[circle, fill=OliveGreen] (2) at (5, 0) {2};
            \node[circle, draw=OliveGreen] (4) at (2, 0) {4};
            \node[rectangle, fill=lime, scale=2] (9) at (1.5, 1.5) {};
            \node[circle, fill=lime] (10) at (2.5, 2) {};
            \node[circle, fill=lightgray] (R) at (2.5, 2.75) {};
            
            \draw (R) -- node[right] {{10}} (10);
            \draw (10) -- node[above] {{{\fontsize{14}{16.8}\selectfont\textsb{9}}}} (9);
            \draw[dotted] (9) -- node[right] {{4}} (4);
            \draw (9) -- node[above left] {{0}} (0);
            \draw (10) -- node[above right] {{2}} (2);
        \end{tikzpicture}} \\
        \bottomrule
    \end{tabular}
}
\label{fig:rename_branches}
}
    \setlength{\tabcolsep}{6pt}
    \rmfamily
    \caption{Recovering a tree from a Phylo2Vec vector: example for \v~$= [0, 2, 2, 5, 2]$. (\textsb{a}) Main algorithm for leaf placement. Initially, we consider a tree with two leaves labelled 0 and 1 and an extra temporary root, which ensures that there are $2j - 1$ entries for any position $j \in \{1, \ldots, n-1\}$. This state correspond to \v = [0]. We then process \v~from left to right. \v[$j$] denotes the branch to be split, yielding a new leaf $j$. At the end of each iteration, a branch labelling step described in (\textsb{b}) is performed. First, branches leading to leaves 0, ..., $n-1$ are labelled 0, ..., $n-1$, respectively. Second, the temporary root is always labelled as $2(n-1)$. Third, for internal branches, the next branch ($n$) to label is the branch of a cherry with the highest label $c_{max}$. We then prune out the leaf $c_{max}$, and repeat the same process for internal branches $n+1, ..., 2(n-1) - 1$. See Figure~\ref{fig:cpxity_to_newick} for more details about complexity.}
    \label{fig:to_newick}
\end{figure}
\setlength{\tabcolsep}{6pt}

The addition of a temporary root in the construction of \v~from a tree and vice versa ensures that there are $2j-1$ branches from which a leaf $j$ can descend from. From these requirements, we can build a unique rooted tree  $\mathcal{T}$ by processing \v~from left to right, where \v[$j$] indicates the branch that will split and yield leaf $j$. Figure~\ref{fig:to_newick} shows a detailed example of this scheme, and other example representations for trees with $n=4$ leaves are shown in Figure~\ref{fig:examples_n4}. We also describe (and prove its existence in the Appendix) an inverse algorithm to convert a tree represented in Newick format as a Phylo2Vec vector in Figure~\ref{fig:to_vector}. To ensure consistency, we also describe in Figure~\ref{fig:rename_branches} an algorithm to label branches based on iterative cherry-picking operations.

\begin{figure}[htbp]
    \centering
    \includegraphics[width=0.8\linewidth,keepaspectratio]{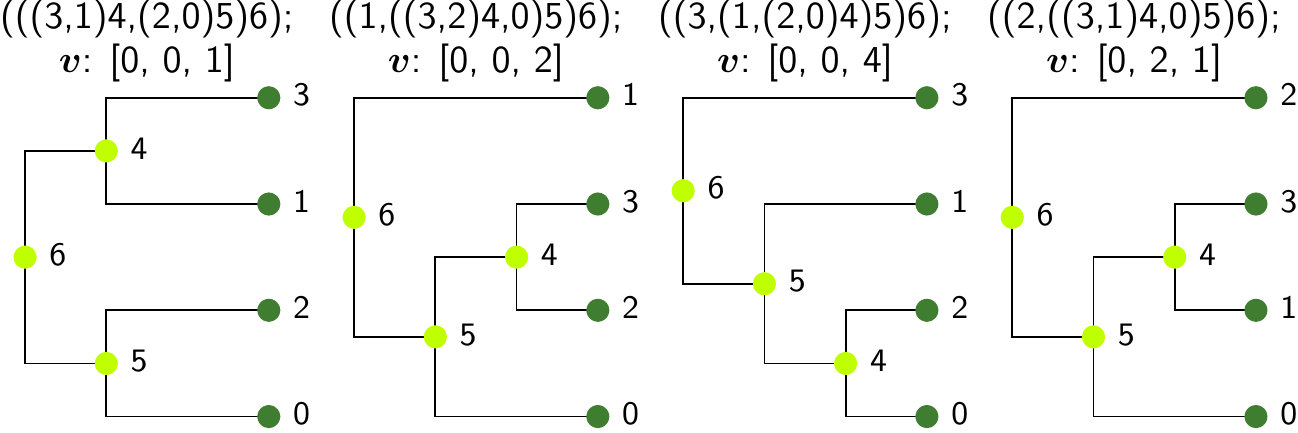}
    \caption{Example of trees with $n=4$ leaves represented in both Newick and Phylo2Vec vector formats. Nodes 0-3 and 4-6 respectively denote the leaves and internal nodes.}
    \label{fig:examples_n4}
    \vspace{0.5em}
    \setlength{\tabcolsep}{8pt}
    \sffamily
    \fontsize{10}{12}\selectfont
    \renewcommand{\arraystretch}{1.43}
    \begin{tabular}{l >{\centering\arraybackslash}b{4.5cm} >{\centering\arraybackslash}b{4.5cm} b{4.9cm}}
         \textsb{Input} & \multicolumn{2}{c}{Newick = (((0,2)4,1)5,3)6;} \\
               & \multicolumn{2}{c}{\begin{tikzpicture}[thick, scale=0.8, every node/.style={scale=0.8}]
            \node[circle, fill=lime] (6) at (1.5, 2) {};
            \node[circle, fill=lime] (4) at (0.5, 1) {};
            \node[circle, fill=lime] (5) at (1, 1.5) {};
            \node[circle, fill=OliveGreen] (0) at (0, 0) {0};
            \node[circle, fill=OliveGreen] (2) at (1, 0) {2};
            \node[circle, fill=OliveGreen] (3) at (3, 0) {3};
            \node[circle, fill=OliveGreen] (1) at (2, 0) {1};
    
            \draw (6) -- node[above left] {{5}} (5);
            \draw (5) -- node[above left] {{4}} (4);
            \draw (6) -- node[above right] {{3}} (3);
            \draw (5) -- node[above right] {{1}} (1);
            \draw (4) -- node[above left] {{0}} (0);
            \draw (4) -- node[above right] {{2}} (2);
        \end{tikzpicture}} \\
         Step 1 & \multicolumn{2}{c}{Next leaf: 1} & \\
               & & \begin{tikzpicture}[thick, scale=0.8, every node/.style={scale=0.8}]
            \node[circle, fill=lightgray] (R) at (1, 2) {};
            \node[circle, fill=lime] (2) at (1, 1) {};
            \node[circle, fill=OliveGreen] (0) at (0, 0) {0};
            \node[circle, fill=orange] (1) at (2, 0) {1};
            
            \draw (R) -- node[right] {{2}} (2);
            \draw (2) -- node[above left] {{0}} (0);
            \draw[red] (2) -- node[above right] {{1}} (1);
        \end{tikzpicture} & Only 1 possible \newline arrangement of a 2-leaf tree \newline $\rightarrow$ \v[1] is always 0. \newline \v~= [\textsb{0}] \\
               &  &  \\
         Step 2 & \multicolumn{2}{c}{Next leaf: 2} &  \\
                & \begin{tikzpicture}[thick, scale=0.8, every node/.style={scale=0.8}]
            \node[circle, fill=lightgray] (R) at (1, 2) {};
            \node[circle, fill=lime] (3) at (1, 1) {};
            \node[circle, fill=OliveGreen] (0) at (0, 0) {0};
            \node[circle, fill=OliveGreen] (1) at (2, 0) {1};
            \node[circle, draw=orange] (2) at (1, 0) {2};
            
            \draw (R) -- node[right] {{2}} (3);
            \draw (3) -- node[above left] {{0}} (0);
            \draw (3) -- node[above right] {{1}} (1);
            \draw[orange, dotted] (0.5, 0.5) -- (2);
        \end{tikzpicture} & \begin{tikzpicture}[thick, scale=0.8, every node/.style={scale=0.8}]
            \node[circle, fill=lightgray] (R) at (1.5, 3) {};
            \node[circle, fill=lime] (6) at (1.5, 2) {};
            \node[circle, fill=lime] (5) at (0.5, 1) {};
            \node[circle, fill=OliveGreen] (0) at (0, 0) {0};
            \node[circle, fill=OliveGreen] (1) at (3,0) {1};
            \node[circle, fill=orange] (2) at (1, 0) {2};

            \draw (6) -- node[above right] {{4}} (R);
            \draw (6) -- node[above left] {{3}} (5);
            \draw (6) -- node[above right] {{1}} (1);
            \draw (5) -- node[above left] {{0}} (0);
            \draw[red] (5) -- node[above right] {{2}} (2);
        \end{tikzpicture}             & Branch 0 splits and yields leaf 2 \newline $\rightarrow$ \v[2] = 0 \newline \v~= [0, \textsb{0}] \\
         Step 3 & \multicolumn{2}{c}{Next leaf: 3} & \\
                & \begin{tikzpicture}[thick, scale=0.8, every node/.style={scale=0.8}]
            \node[circle, fill=lightgray] (R) at (1.5, 3) {};
            \node[circle, fill=lime] (6) at (1.5, 2) {};
            \node[circle, fill=lime] (5) at (0.5, 1) {};
            \node[circle, fill=OliveGreen] (0) at (0, 0) {0};
            \node[circle, fill=OliveGreen] (2) at (1, 0) {2};
            \node[circle, fill=OliveGreen] (1) at (3,0) {1};
            \node[circle, draw=orange] (3) at (4,0) {3};

            \draw (6) -- node[above right] {{4}} (R);
            \draw (6) -- node[above left] {{3}} (5);
            \draw (6) -- node[above right] {{1}} (1);
            \draw (5) -- node[above left] {{0}} (0);
            \draw (5) -- node[above right] {{2}} (2);
            \draw[orange, dotted] (3) -- (1.5, 2.5);
        \end{tikzpicture} & \begin{tikzpicture}[thick, scale=0.8, every node/.style={scale=0.8}]
            \node[circle, fill=lime] (6) at (1.5, 2) {};
            \node[circle, fill=lime] (4) at (0.5, 1) {};
            \node[circle, fill=lime] (5) at (1, 1.5) {};
            \node[circle, fill=OliveGreen] (0) at (0, 0) {0};
            \node[circle, fill=OliveGreen] (2) at (1, 0) {2};
            \node[circle, fill=orange] (3) at (3, 0) {3};
            \node[circle, fill=OliveGreen] (1) at (2, 0) {1};
    
            \draw (6) -- node[above left] {{5}} (5);
            \draw (5) -- node[above left] {{4}} (4);
            \draw (6) -- node[above right] {{3}} (3);
            \draw (5) -- node[above right] {{1}} (1);
            \draw (4) -- node[above left] {{0}} (0);
            \draw (4) -- node[above right] {{2}} (2);
        \end{tikzpicture}             & Branch 4 splits and yields leaf 3 \newline $\rightarrow$ \v[3] = 4 \newline \v~= [0, 0, \textsb{4}] \\
    \end{tabular}
    \setlength{\tabcolsep}{6pt}
    \rmfamily
    \caption{Labelling a tree as a Phylo2Vec vector \v: example for \v~= [0, 0, 4]. We process leaves in ascending order. For each leaf $j$, we determine the branch that split and yielded leaf $j$, which corresponds to \v$[j]$. At each step, we re-label the branches with the same process as in Figure 2.
    Figure 5: Comparison of Phylo2Vec moves with three popular tree distances: subtree-prune-and-regraft (SPR; left), Robinson-Foulds (RF; middle), and Kuhner-Felsenstein (KF; right). To generate the distances, a random walk of 5000 steps was performed from a random initial \v~with 200 taxa. At each step, each $v_i$ can increment, decrement or remain unchanged.}
    \label{fig:to_vector}
\end{figure}

\subsubsection*{Complexity}
The algorithm underlying Figure~\ref{fig:to_newick} and detailed in Algorithm~\ref{alg:to_newick} has an average time complexity of $\mathcal{O}(n^{1.5})$ and a worst-case time complexity of $\mathcal{O}(n^2)$ due to the linear cost insert step. However, an implementation using a data structure based on Adelson-Velsky and Landis (AVL) trees is possible which would allow Algorithm~\ref{alg:to_newick} to run in linearithmic time ($\mathcal{O}(n \log n)$; an implementation is available at~\citep{scheidwasser2024}). This self-balancing tree implementation is only faster for large trees (over $\approx 50000$ taxa). A basic version using \texttt{NumPy}~\citep{harris2020} runs in a few milliseconds for $n=1000$ taxa on a modern CPU. The inverse algorithm (converting a Newick string to a Phylo2Vec \v), detailed in Algorithm~\ref{alg:to_vector}, is also of linearithmic complexity when internal nodes are already labelled (according to the scheme described in Fig.~\ref{fig:to_newick}) and quadratic otherwise (see Fig.~\ref{fig:cpxity_to_vector}). Speedups are made through just-in-time compilation, e.g., using \texttt{Numba}~\citep{lam2015} in Python.


\subsubsection*{Distances between trees} 
The formulation of Phylo2Vec as a one-to-one correspondence between binary trees and integer vectors constrained by Equation~\ref{eq:phylo2vec} naturally allows for a new measure of distance between trees. For any two Phylo2Vec trees \v~and $\boldsymbol{w}$, a Hamming distance can be defined as

\begin{equation}
    \mu(\text{\v}, \boldsymbol{w}) = \sum_{i=1}^{n-1} \mathbb{I}_{v_i \neq w_i}
\end{equation}

To compare this distance with other tree distance metrics, we consider a simple discrete random walk in the space of possible Phylo2Vec vectors $\mathbb{V}$. At each step, we create a new vector $\boldsymbol{w}$ from the previous vector \v~as follows. First, we choose a random subset of the indices $\mathcal{I} \subseteq \{2\ldots,n-2\}$. For each $i \notin \mathcal{I}$, we set $w_i =v_i$ and for each $i \in \mathcal{I}$, $w_i = \min(2(i-1),\allowbreak\max(v_i+J(i),0))$ where the $J(i)$ are iid random variables uniform on the set $\{-1,1\}$. Note that the values of $J(i)$ at different steps of the walk are also independent, and that the minimum and maximum in the definition of $w_i$ ensure that it satisfies the constraint $0 \leq w_i \leq 2(i-1)$.

\begin{figure}[htbp]
    \centering
    \includegraphics[width=\linewidth]{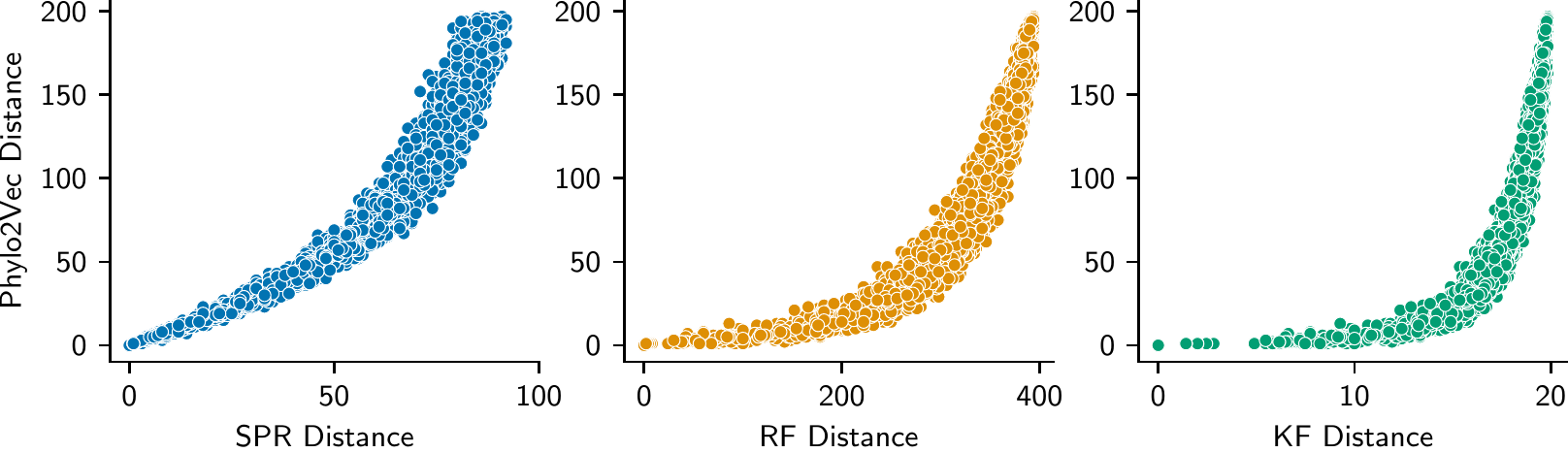}
    \caption{Comparison of Phylo2Vec moves with three popular tree distances: subtree-prune-and-regraft (SPR; left), Robinson-Foulds (RF; middle), and Kuhner-Felsenstein (KF; right). To generate the distances, a random walk of 5000 steps was performed from a random initial \v~with 200 taxa. At each step, each $v_i$ can increment, decrement or remain unchanged.}
    \label{fig:distance}
\end{figure}

As $1,n-1 \notin \mathcal{I}$, we fix $w_1 = 0$ (by our constraints) and $w_{n-1}=2(n-2)$ (to ensure that we move in the unrooted tree space for SPR distance calculations). Figure~\ref{fig:distance} compares $\mu$ to an approximate SPR distance~\citep{deoliveira2016}, Robinson-Foulds (RF) distance~\citep{robinson1981} and Kuhner-Felsenstein (KF) distance~\citep{kuhner1994}. We note that exact, rooted distance for SPR is nondeterministic polynomial time (NP)-hard to compute~\citep{Bordewich2005-ix} and therefore cannot be directly compared to our rooted Phylo2Vec formulation. For all distances, we see a nonlinear correspondence, especially for RF and KF distance. Small changes in \v~can lead to very large topological jumps, but equally, small jumps are also possible. Modifying several indexes in \v~results in significant jumps across tree space, leading to new trees that are very dissimilar. As a result, SPR, RF, or KF distances saturate as we increase the number of changes in \v~\citep{St_John2017-vq}. However, we note that small changes in \v$_i$ can also readily correspond to very minor topological changes. 

In the exploration of tree space, the number of possible moves for both SPR and Phylo2Vec is of order $\mathcal{O}(n^2)$ (see~\nameref{sec:phylo2vec} in the Appendix). Consequently, Phylo2Vec is expected to explore tree space in a similar manner than SPR, with proposals being less local than nearest neighbour interchange (NNI) but also less global than those by tree bisection and reconnection (TBR). However, the number of \emph{single} SPR changes is approximately four times greater than the number of \emph{single} changes in \v~(that is, changes of a single index of \v)~as SPR considers internal nodes while \v~is defined across the leaves, and so Phylo2Vec changes are likely a subset of possible SPR changes.

Whereas Figure~\ref{fig:distance} shows distances between unrooted trees, our framework is built on rooted phylogenies at its core. Knowing that all rootings produce the same likelihood due to the pulley principle and reversibility of nucleotide substitution models~\citep{felsenstein2004}, we can, for any rooted phylogeny, switch to one that is rooted at a different outgroup and has exactly the same likelihood. Thus, an equivalence class $\mathcal{V}$ exists where, given a likelihood or parsimony score $\ell$, any given Phylo2Vec vector \v $~\in \mathcal{V}$ has the same $\ell(\text{\v})$,~an SPR or RF distance of 0, but a Phylo2Vec distance of $\mu>0$. In practice, $\mu$ is often very large between $\text{\v}\in\mathcal{V}$ (comparable to half the maximum SPR distance, see Fig.~\ref{fig:equiv_class}), which makes switching a vector \v~$\in \mathcal{V}$ to an equivalent \v$' \in \mathcal{V}$ an additional mechanism for tree space exploration.

\subsubsection*{Shuffling Indices}
This distance between two trees is not symmetric with respect to the labelling of the trees, as discussed further in the Appendix (\nameref{sec:phylo2vec}). Depending on the choice of labelling, certain portions of the tree may be easier to optimise than others when performing phylogenetic inference. This is an undesirable quality and can be remedied by a simple reordering of indices within our algorithm. An example of a reordering algorithm is presented in Figure~\ref{fig:reorder_v2}.

\begin{figure}
\small
\sffamily
\begin{tabular}{l m{5cm} c m{5cm}}
    Input & Newick: ((((0,2),5),1),(3,(4,6))); \\
    & \v: [0, 0, 4, 3, 6, 4] \\
    \midrule
    Initial tree & \begin{tikzpicture}[thick, scale=.75, every node/.style={scale=0.9}]
                        \node[circle, fill=OliveGreen] (0) at (0, 0) {0};
                        \node[circle, fill=OliveGreen] (2) at (1, 0) {2};
                        \node[circle, fill=OliveGreen] (5) at (2, 0) {5};
                        \node[circle, fill=OliveGreen] (1) at (3, 0) {1};
                        \node[circle, fill=OliveGreen] (3) at (4, 0) {3};
                        \node[circle, fill=OliveGreen] (4) at (5, 0) {4};
                        \node[circle, fill=OliveGreen] (6) at (6, 0) {6};
                        \node[circle, fill=lime] (9) at (0.5, 1) {};
                        \node[circle, fill=lime] (10) at (1.25, 2) {};
                        \node[circle, fill=lime] (11) at (2.125, 3) {};
                        \node[circle, fill=lime] (7) at (5.5, 2) {};
                        \node[circle, fill=lime] (8) at (4.75, 3) {};
                        \node[circle, fill=lime] (12) at (3.4375, 4) {};
                        \node[circle, fill=none, style={font=\vphantom{Ag}}] (a) at (0, -0.75) {a};
                        \node[circle, fill=none, style={font=\vphantom{Ag}}] (c) at (1, -0.75) {c};
                        \node[circle, fill=none, style={font=\vphantom{Ag}}] (f) at (2, -0.75) {f};
                        \node[circle, fill=none, style={font=\vphantom{Ag}}] (b) at (3, -0.75) {b};
                        \node[circle, fill=none, style={font=\vphantom{Ag}}] (d) at (4, -0.75) {d};
                        \node[circle, fill=none, style={font=\vphantom{Ag}}] (e) at (5, -0.75) {e};
                        \node[circle, fill=none, style={font=\vphantom{Ag}}] (g) at (6, -0.75) {g};
                    
                        \draw (9) -- (0);
                        \draw (9) -- (2);
                        \draw (10) -- (9);
                        \draw (10) -- (5);
                        \draw (11) -- (10);
                        \draw (11) -- (1);
                        \draw (7) -- (4);
                        \draw (7) -- (6);
                        \draw (8) -- (7);
                        \draw (8) -- (3);
                        \draw (12) -- (11);
                        \draw (12) -- (8);
                     \end{tikzpicture} & Level 2 & \begin{tikzpicture}[thick, scale=.75, every node/.style={scale=0.9}]
                        \node[circle, fill=OliveGreen] (0) at (0, 0) {};
                        \node[circle, fill=OliveGreen] (2) at (1, 0) {};
                        \node[circle, fill=orange] (5) at (2, 0) {2};
                        \node[circle, fill=OliveGreen] (1) at (3, 0) {0};
                        \node[circle, fill=OliveGreen] (3) at (4, 0) {1};
                        \node[circle, fill=orange] (4) at (5, 0) {3};
                        \node[circle, fill=orange] (6) at (6, 0) {4};
                        \node[circle, fill=lime] (9) at (0.5, 1) {};
                        \node[circle, fill=Apricot] (10) at (1.25, 2) {};
                        \node[circle, fill=lime] (11) at (2.125, 3) {};
                        \node[circle, fill=Apricot] (7) at (5.5, 2) {};
                        \node[circle, fill=lime] (8) at (4.75, 3) {};
                        \node[circle, fill=lime] (12) at (3.4375, 4) {};
                    
                        \draw (9) -- (0);
                        \draw (9) -- (2);
                        \draw (10) -- (9);
                        \draw (10) -- (5);
                        \draw (11) -- (10);
                        \draw (11) -- (1);
                        \draw (7) -- (4);
                        \draw (7) -- (6);
                        \draw (8) -- (7);
                        \draw (8) -- (3);
                        \draw (12) -- (11);
                        \draw (12) -- (8);
                     \end{tikzpicture}\\
        Level 1 & \begin{tikzpicture}[thick, scale=.75, every node/.style={scale=0.9}]
                        \node[circle, fill=OliveGreen] (0) at (0, 0) {};
                        \node[circle, fill=OliveGreen] (2) at (1, 0) {};
                        \node[circle, fill=OliveGreen] (5) at (2, 0) {};
                        \node[circle, fill=orange] (1) at (3, 0) {0};
                        \node[circle, fill=orange] (3) at (4, 0) {1};
                        \node[circle, fill=OliveGreen] (4) at (5, 0) {};
                        \node[circle, fill=OliveGreen] (6) at (6, 0) {};
                        \node[circle, fill=lime] (9) at (0.5, 1) {};
                        \node[circle, fill=lime] (10) at (1.25, 2) {};
                        \node[circle, fill=Apricot] (11) at (2.125, 3) {};
                        \node[circle, fill=lime] (7) at (5.5, 2) {};
                        \node[circle, fill=Apricot] (8) at (4.75, 3) {};
                        \node[circle, fill=lime] (12) at (3.4375, 4) {};
                    
                        \draw (9) -- (0);
                        \draw (9) -- (2);
                        \draw (10) -- (9);
                        \draw (10) -- (5);
                        \draw (11) -- (10);
                        \draw (11) -- (1);
                        \draw (7) -- (4);
                        \draw (7) -- (6);
                        \draw (8) -- (7);
                        \draw (8) -- (3);
                        \draw (12) -- (11);
                        \draw (12) -- (8);
                     \end{tikzpicture} & Level 3& \begin{tikzpicture}[thick, scale=.75, every node/.style={scale=0.9}]
                    \node[circle, fill=orange] (0) at (0, 0) {5};
                    \node[circle, fill=orange] (2) at (1, 0) {6};
                    \node[circle, fill=OliveGreen] (5) at (2, 0) {2};
                    \node[circle, fill=OliveGreen] (1) at (3, 0) {0};
                    \node[circle, fill=OliveGreen] (3) at (4, 0) {1};
                    \node[circle, fill=OliveGreen] (4) at (5, 0) {3};
                    \node[circle, fill=OliveGreen] (6) at (6, 0) {4};
                    \node[circle, fill=Apricot] (9) at (0.5, 1) {};
                    \node[circle, fill=lime] (10) at (1.25, 2) {};
                    \node[circle, fill=lime] (11) at (2.125, 3) {};
                    \node[circle, fill=lime] (7) at (5.5, 2) {};
                    \node[circle, fill=lime] (8) at (4.75, 3) {};
                    \node[circle, fill=lime] (12) at (3.4375, 4) {};
                
                    \draw (9) -- (0);
                    \draw (9) -- (2);
                    \draw (10) -- (9);
                    \draw (10) -- (5);
                    \draw (11) -- (10);
                    \draw (11) -- (1);
                    \draw (7) -- (4);
                    \draw (7) -- (6);
                    \draw (8) -- (7);
                    \draw (8) -- (3);
                    \draw (12) -- (11);
                    \draw (12) -- (8);
                 \end{tikzpicture} \\
    \midrule
    Output & Newick: ((((5,6),2),0),(1,(3,4))); \\
    & \v: [0, 0, 1, 3, 2, 5] \\
    & \begin{tikzpicture}[thick, scale=.75, every node/.style={scale=0.9}]
                    \node[circle, fill=OliveGreen] (0) at (0, 0) {5};
                    \node[circle, fill=OliveGreen] (2) at (1, 0) {6};
                    \node[circle, fill=OliveGreen] (5) at (2, 0) {2};
                    \node[circle, fill=OliveGreen] (1) at (3, 0) {0};
                    \node[circle, fill=OliveGreen] (3) at (4, 0) {1};
                    \node[circle, fill=OliveGreen] (4) at (5, 0) {3};
                    \node[circle, fill=OliveGreen] (6) at (6, 0) {4};
                    \node[circle, fill=lime] (9) at (0.5, 1) {};
                    \node[circle, fill=lime] (10) at (1.25, 2) {};
                    \node[circle, fill=lime] (11) at (2.125, 3) {};
                    \node[circle, fill=lime] (7) at (5.5, 2) {};
                    \node[circle, fill=lime] (8) at (4.75, 3) {};
                    \node[circle, fill=lime] (12) at (3.4375, 4) {};
                    \node[circle, fill=none, style={font=\vphantom{Ag}}] (a) at (0, -0.75) {a};
                    \node[circle, fill=none, style={font=\vphantom{Ag}}] (c) at (1, -0.75) {c};
                    \node[circle, fill=none, style={font=\vphantom{Ag}}] (f) at (2, -0.75) {f};
                    \node[circle, fill=none, style={font=\vphantom{Ag}}] (b) at (3, -0.75) {b};
                    \node[circle, fill=none, style={font=\vphantom{Ag}}] (d) at (4, -0.75) {d};
                    \node[circle, fill=none, style={font=\vphantom{Ag}}] (e) at (5, -0.75) {e};
                    \node[circle, fill=none, style={font=\vphantom{Ag}}] (g) at (6, -0.75) {g};
                
                    \draw (9) -- (0);
                    \draw (9) -- (2);
                    \draw (10) -- (9);
                    \draw (10) -- (5);
                    \draw (11) -- (10);
                    \draw (11) -- (1);
                    \draw (7) -- (4);
                    \draw (7) -- (6);
                    \draw (8) -- (7);
                    \draw (8) -- (3);
                    \draw (12) -- (11);
                    \draw (12) -- (8);
                 \end{tikzpicture} \\
\end{tabular}
\rmfamily
\caption{Example of a reordering scheme of \v~using level-order traversal. Starting from the root, for each level, we relabel the immediately descending leaf nodes with the smallest integers available (from 0 to $n-1$). The letters (a-g) indicate the taxa, showing that reordering the leaves does not affect tree topology but simply changes the integer-taxon mapping.}
\label{fig:reorder_v2}
\end{figure}

Consider a tree $\mathcal{T}$ where the leaves are labelled by a fixed set of indices $\{1,2,\ldots,n-1\}$. Suppose that $\sigma$ is a permutation of $\{1,2,\ldots,n-1\}$, and consider a shuffled tree $\sigma(\mathcal{T})$ with the same topological structure as $\mathcal{T}$, but where, for each $j \in \{1,2,\ldots,n-1\}$, the leaf with original label $i$ now has label $\sigma(j)$.

Calculating the likelihood requires a tree $\mathcal{T}$ and a set of genetic data $\mathcal{D} = (D_1,D_2,\ldots,D_{n-1})$, where $D_j$ corresponds to the physical or genetic characteristics of leaf $j$. We can then write the likelihood as $L = L(\mathcal{T},\mathcal{D})$. Moreover, defining the shuffled genetic data as $\sigma(D) = (D_{\sigma^{-1}(1)},D_{\sigma^{-1}(2)},\ldots,D_{\sigma^{-1}(n-1)})$, we can then see that $ L(\mathcal{T},\mathcal{D}) = L(\sigma(\mathcal{T}),\sigma(\mathcal{D}))$. This occurs because when computing the likelihood, any calculation for $L(\mathcal{T},\mathcal{D})$ that involves the node with original label $i$ (and hence genetic data $D_i$) will now involve the node with label $\sigma(i)$ and hence genetic data $D_{\sigma^{-1}(\sigma(i))} = D_i$. Should the permutation only be applied to either the tree labels or the genetic data set, the resulting likelihood will likely be different from $ L(\mathcal{T},\mathcal{D})$. Thus, since the topological structure of $\mathcal{T}$ is the same as $D(\mathcal{T})$, the likelihood will remain unchanged. A more rigorous proof can be found in the Appendix (\nameref{sec:phylo2vec}).

One can also recover the vector $\boldsymbol{v}$ corresponding to the shuffled tree $\sigma(T)$. This is possible because of the bijective relationship between the space of $\boldsymbol{v}$'s and the space of trees. We provide an algorithm that inverts our map from $\boldsymbol{v}$ to $M$ in the Appendix (\nameref{sec:phylo2vec}). Thus, one can equivalently define a shuffled vector $\sigma(\boldsymbol{v})$ (such that $\sigma(\boldsymbol{v})$ generates $\sigma(\mathcal{T})$) and consider the likelihood relationship as $L(\boldsymbol{v},\mathcal{D}) = L(\sigma(\boldsymbol{v}),\sigma(\mathcal{D}))$. This allows for discrete optimisation steps to be taken with respect to the new shuffled $\boldsymbol{v}$, increasing the flexibility of the algorithm while removing the asymmetric effects of the initial labelling.

\subsubsection*{Branch lengths}
In addition to tree topology, determining the branch lengths of a tree is an important facet in phylogenetic inference. When making small changes to the tree topology, a number of portions of the tree will remain identical. Therefore, it is likely that the optimal values of subtree branch lengths will not change. It is therefore helpful to represent branch lengths in a method that is robust to these changes to avoid carrying out the full optimisation process every time the topology is changed.

Within the Phylo2Vec framework, there are several approaches in which branch lengths can be integrated. First, given each \v$_j$ refers to the branch splitting and leading to leaf $j$, a simple solution would consist in adding a 2-column matrix specifying the position at which branch \v$_j$ splits and the length of the new branch that produces leaf $j$. Alternatively, it is possible to assign each leaf node a ``position'', calculate internal node positions as some weighted average of the positions of the nearby leaf nodes, and then calculate branch lengths based on the distance between a pair of nodes. This would have the advantage of branch lengths being independent of the choice of root, thus allowing to easily switch between the unrooted equivalence classes discussed previously. 

For the examples in this paper, we used \texttt{RAxML-NG}~\citep{kozlov2019} to optimise the branch lengths at each step of the algorithm without using information from previous branch lengths. This reduces the speed of the optimisation and is an area for improvement in future work.

\subsection{Evaluation}\label{sec:optim}
\subsubsection{Problem and data}
To demonstrate the utility of Phylo2Vec, we apply our new representation for phylogenetic inference of five popular empirical molecular sequence datasets under the maximum likelihood (ML) criterion. This dataset corpus spans across different biological entities, taxa, and genetic sequence sizes. 

\begin{table}[htbp]
\centering
\caption{Evaluation datasets, sorted by number of taxa.}
\begin{tabular}{llllr} 
\toprule
Name     & Reference & Type      & \# taxa & \# bases \\
\midrule
Yeast    & \citet{rokas2003, schliep2011} & Fungi     & 8       & 127,018  \\
H3N2     & \citet{sagulenko2018} & Virus       & 19      & 1,407    \\
M501     & \citet{Garey1996-jc} & Animal      & 29      & 2,520    \\
FluA     & \citet{fourment2019} & Virus       & 69      & 987 \\
Zika     & \citet{sagulenko2018} & Virus       & 86      & 10,807    \\
\bottomrule
\end{tabular}
\label{tab:data}
\end{table}


It has been proved that ML inference for phylogenetic trees is NP-hard~\citep{Roch2006-qs} and therefore our key goal is to define a sensible heuristic that can explore the vastness of tree space. Moreover, the likelihood surface exhibits high curvature~\citep{sanderson2011} and being trapped in a local optima is a persistent problem across all heuristic phylogenetic approaches. 

\subsubsection{Tree topology optimisation using hill-climbing}

A simple way to explore the space of possible trees is to use hill climbing where we simply compute the difference in likelihood after a single element is changed. We define the neighbour matrix
\begin{equation}
    \Delta \ell(\text{\v})_{ij} = \ell((v_1,\ldots,v_{i-1},j,v_{i+1},\ldots,v_{n-1})) - \ell(\text{\v}),
\end{equation}
that is, the tree considered in the first likelihood has identical entries except for the $i^{\text{th}}$ entry, which is changed to $j$. For $(i,j)$ such that $v_i = j$ is infeasible, we set $\Delta \ell_{ij} = 0$. We have found that considering each row of the neighbour matrix yields good results, i.e., if $\max(\Delta \ell_i) > 0$, then we find $j = \text{argmax}_j(\Delta \ell_{ij})$ and change the value of $v_i$ to $j$. This algorithm is guaranteed to converge to a point where $\max(\Delta \ell) \leq 0$ as no change in $\boldsymbol{v}$ results in a gradient that is greater than zero. Moreover, as there are only finitely many possible $\boldsymbol{v}$, and $\ell$ is strictly decreasing after each iteration of the while loop, the algorithm must converge in finite time. More complicated optimisation algorithms can be readily created and is an especially useful aspect of our representation. An example is performing hill-climbing over paired changes in \v. Exploratory analysis suggests that paired changes are far more robust to being trapped in local minima, but at the cost of higher complexity. For challenging phylogenies, a simpler parsimony or minimum evolution score can be used to perform hill-climbing over pairs as an exploratory search.



However, as highlighted above, a fundamental asymmetry exists in Phylo2Vec which can make optimisation inefficient. A simple solution to mitigating this asymmetry is to reorder the integer-taxon mapping to obtain an ordered vector (and thus, an ordered tree), as described previously in~\nameref{sec:ordered} and Figure~\ref{fig:reorder_v2}. The advantage of carrying out our hill climbing scheme on these ordered trees is that it removes the secondary effects of changing an element of \v~which can occur by the divergence in internal node labels. This prevents our algorithm from getting stuck in local minima, as it means that more parts of the tree can be easily edited.  


The resulting algorithm is detailed in Algorithm~\ref{alg:hill_climbing}. Our investigations have shown that all the possible trees that are one step from some ordered \v~are also one SPR move from the original tree (though the converse is not true - not all SPR moves will be one step from \v). This is proved in the Appendix (\nameref{sec:phylo2vec}). Thus, this application of our Phylo2Vec formulation falls within the SPR framework and provides a mathematically convenient and principled way to explore tree space using well-tested SPR methodology.

We note that we could additionally explore rooted equivalence classes to further prevent being stuck in local minima. In particular, there is more freedom in the movements of nodes further down the tree, and rerooting at the deepest node would allow all nodes to be easily moved to a variety of locations. However, for the experiments presented hereinafter, we found this extra degree of freedom to be unnecessary.

\begin{algorithm}[htbp]
\caption{Hill-climbing optimisation of a tree with $n$ leaves}
\label{alg:hill_climbing}
\begin{algorithmic}
\State \textbf{Input} $\text{\v}\in \mathcal{T}_n$ \Comment{Initialise with a random \v}
\State $\ell_{\text{best}} \gets \ell(\text{\v})$ \Comment{Initial best likelihood value}
\Repeat
\State $\text{Reorder}(\text{\v})$ \Comment{Reorder the labels (see Fig.~\ref{fig:reorder_v2})}
\State Sample $i\in \{1,\ldots, n-1\}$ \Comment{Sample an index of \v}
\State $\boldsymbol{G_i} \gets \Delta\ell(\text{\v})_{ij} $ \Comment{$G_{ij}$ = likelihood difference from changing $v_i$ to j}
\State $j \gets \text{argmax}(\boldsymbol{G_i})$
\Comment{Find the best change}
\State $v_i \rightarrow j$ \Comment{Change $v_i$}
\State $\ell_{\text{best}} \gets \ell(\text{\v})$
\Until{$\max(G_{i=1, \ldots, n-1}) = 0$} \Comment{Continue iterating until local minimum}
\end{algorithmic}
\end{algorithm}

\subsubsection*{Additional properties of the Phylo2Vec vector}

\begin{figure}[t!]
    \centering
    \subfloat[]{\includegraphics[width=0.48\linewidth]{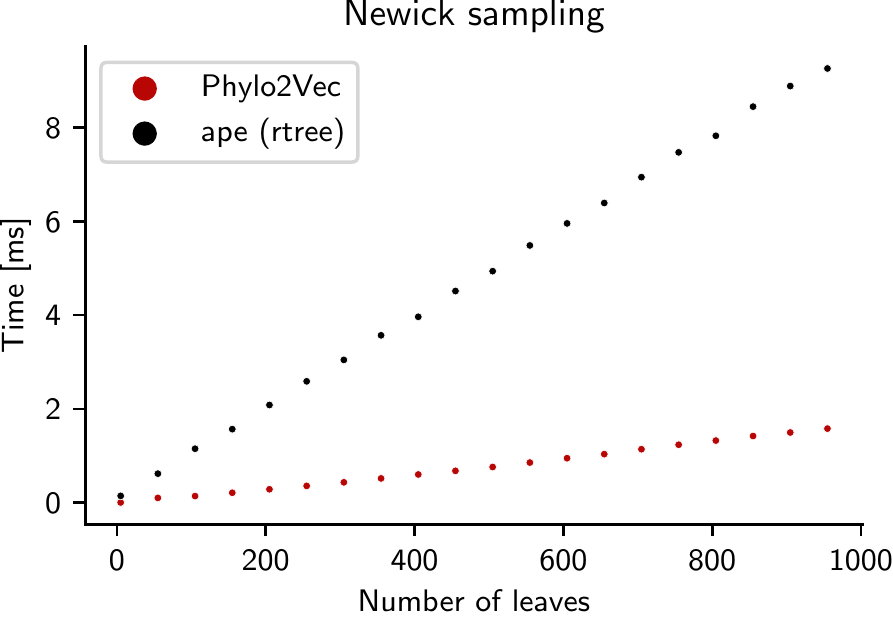} \label{fig:sampling_time}}
    \subfloat[]{\includegraphics[width=0.48\linewidth]{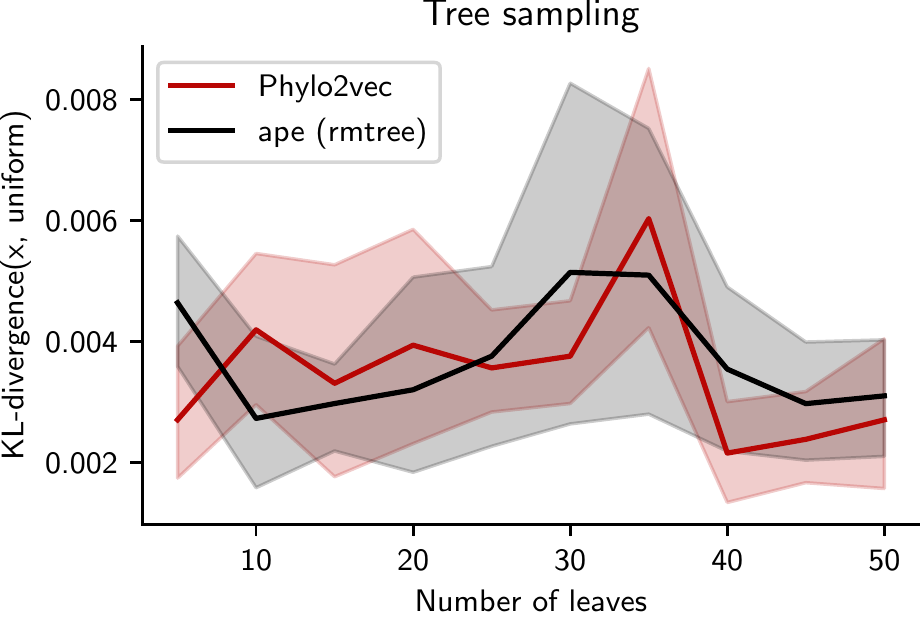} \label{fig:sampling_div}}
    \\
    \subfloat[]{\includegraphics[width=0.48\linewidth]{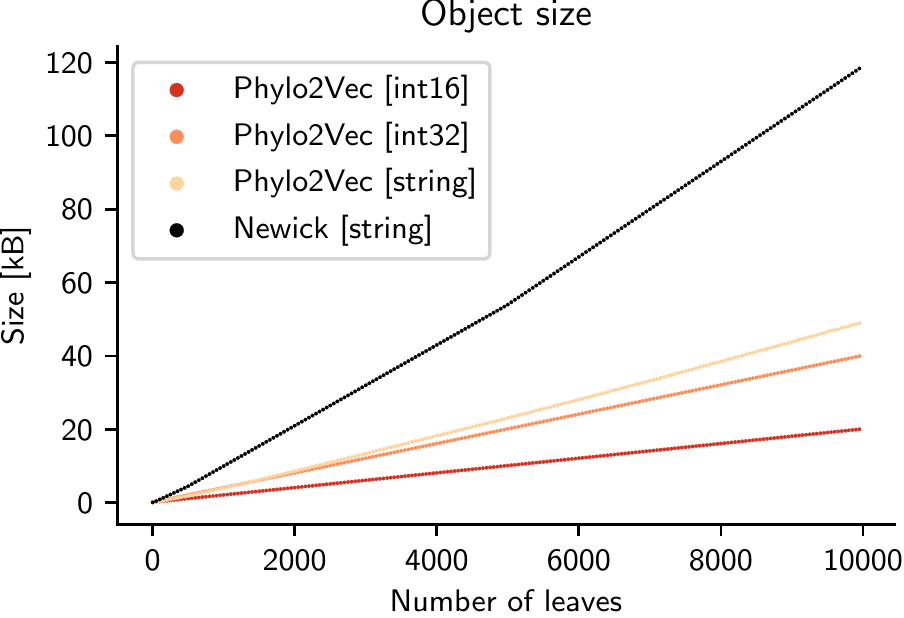} \label{fig:sizes}}
    \subfloat[]{\includegraphics[width=0.48\linewidth]{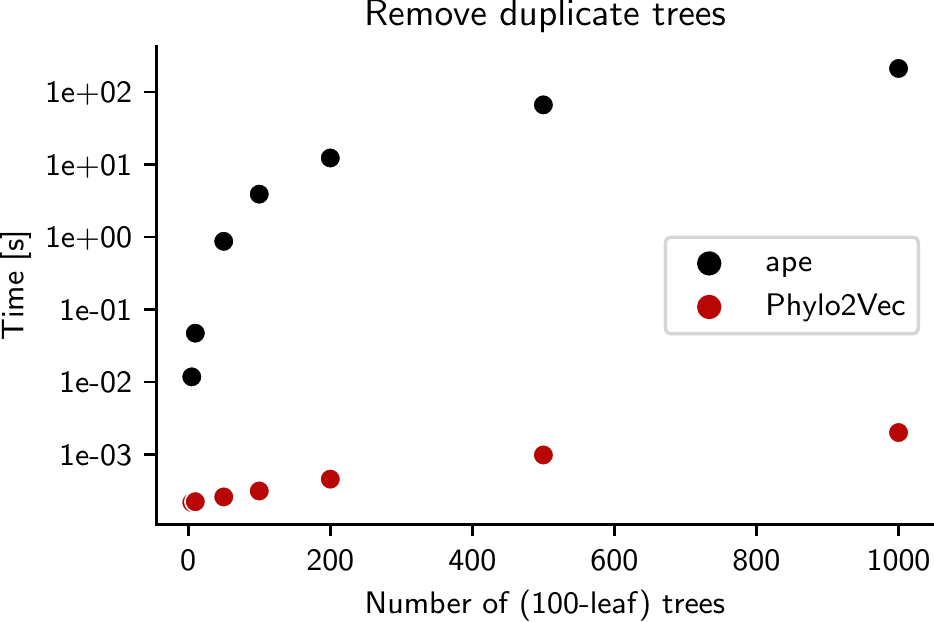} \label{fig:unique_times}}
    \caption{Phylo2Vec allows for fast and unbiased sampling, low memory or storage, and fast comparison of trees. \textsb{(a)} Average sampling time using \texttt{phylo2vec.utils.sample} and \texttt{rtree} from \texttt{ape}. Execution time was measured over 100 executions using Python's \texttt{timeit} and R's \texttt{microbenchmark}, respectively. \textsb{(b)} Sampling bias comparison. For each size and sampler, we sample 10000 trees and converted them first to their Phylo2Vec representation, and second to an integer using a method similar to that of~\citet{rohlf1983}. We then compare the probability distributions of the integers generated by Phylo2Vec and \texttt{ape} sampling against the reference uniform distribution for each tree size using the Kullback-Leibler (KL) divergence. The lower the KL-divergence value, the more the reference distribution and the distribution of interest share similar information. \textsb{(c)} Object sizes for different tree sizes of Phylo2Vec vectors (stored as a 16- or 32-bit \texttt{numpy} integer array, or a string) compared against their Newick-format equivalents (without branch length information). \textsb{(d)} Average time for duplicate removal from a set of trees using Phylo2Vec (vectors) and the \textit{unique.multiPhylo} function from \texttt{ape}. Execution time was measured over 30 executions using Python's \texttt{timeit} and R's \texttt{microbenchmark}, respectively.}
    \label{fig:benchmarks}
\end{figure}

An additional advantage of having an integer vector representation for binary trees such as Phylo2vec is efficiency with respect to sampling, data storage, as well as assessing tree equality (with respect to topology). We highlight these properties in Figure~\ref{fig:benchmarks} by performing several benchmarks against functions of shows the widely used R library \texttt{ape}~\cite{paradis2019}. Figure~\ref{fig:sampling_time} shows how Phylo2Vec sampling of trees is several times faster than the function \texttt{rtree}, while also being simple in construction and implementation. Figure~\ref{fig:sampling_div} verifies that the Phylo2Vec sampling distribution is indeed uniform. While we do explore other sampling schemes further, ordered trees present one avenue to perform constrained tree sampling. Figure~\ref{fig:sizes} shows the storage costs in kB of Phylo2Vec as compared to a Newick string with \emph{only} topological information. From these simple simulations we estimate a Phylo2Vec vector can be stored as an integer array or a string as much as a six times reduced storage cost. Finally, Figure~\ref{fig:unique_times} shows the time required to find a unique set of topologies from a set of trees. Phylo2Vec is several orders of magnitude faster than \texttt{unique.multiPhylo} in \texttt{ape}, and can be massively parallelised. This speed difference can be particularly useful in Bayesian settings.

\subsection{Implementation}
All Phylo2Vec algorithms and related optimisation methods presented in the main text were implemented in Python 3.10 using \texttt{NumPy}~\citep{harris2020} and \texttt{numba}~\citep{lam2015}. Tree manipulation scripts were written using \texttt{ete3}~\citep{huerta2016}. Time complexity was estimated manually and verified empirically using \texttt{big\_o}~(\url{https://github.com/pberkes/big_O}). Dataset construction was based on \texttt{phangorn}~\citep{schliep2011} in R and \texttt{TreeTime}~\citep{sagulenko2018} in Python. Maximum likelihood estimation was performed using \texttt{RAxML-NG}~\citep{kozlov2019}. An implementation is available at: \url{https://github.com/Neclow/phylo2vec}. Execution times were benchmarked using Python's \texttt{timeit} on a machine equipped with a 64-core CPU @ 7 GHz, with 256 GB of RAM.

\section{Results}
We test Phylo2Vec by performing inference on five popular empirical datasets described in Table~\ref{tab:data}. This dataset corpus spans across different biological entities, taxa, and genetic sequence sizes. 

For each dataset, we use the optimisation procedure described in~\nameref{sec:optim}, using \texttt{RAxML-NG} for branch length and substitution matrix optimisation. We report performance using the negative log version of the tree likelihood defined by~\citet{felsenstein1983}.

\begin{figure}[htbp]
    \centering
    \includegraphics[width=0.75\linewidth]{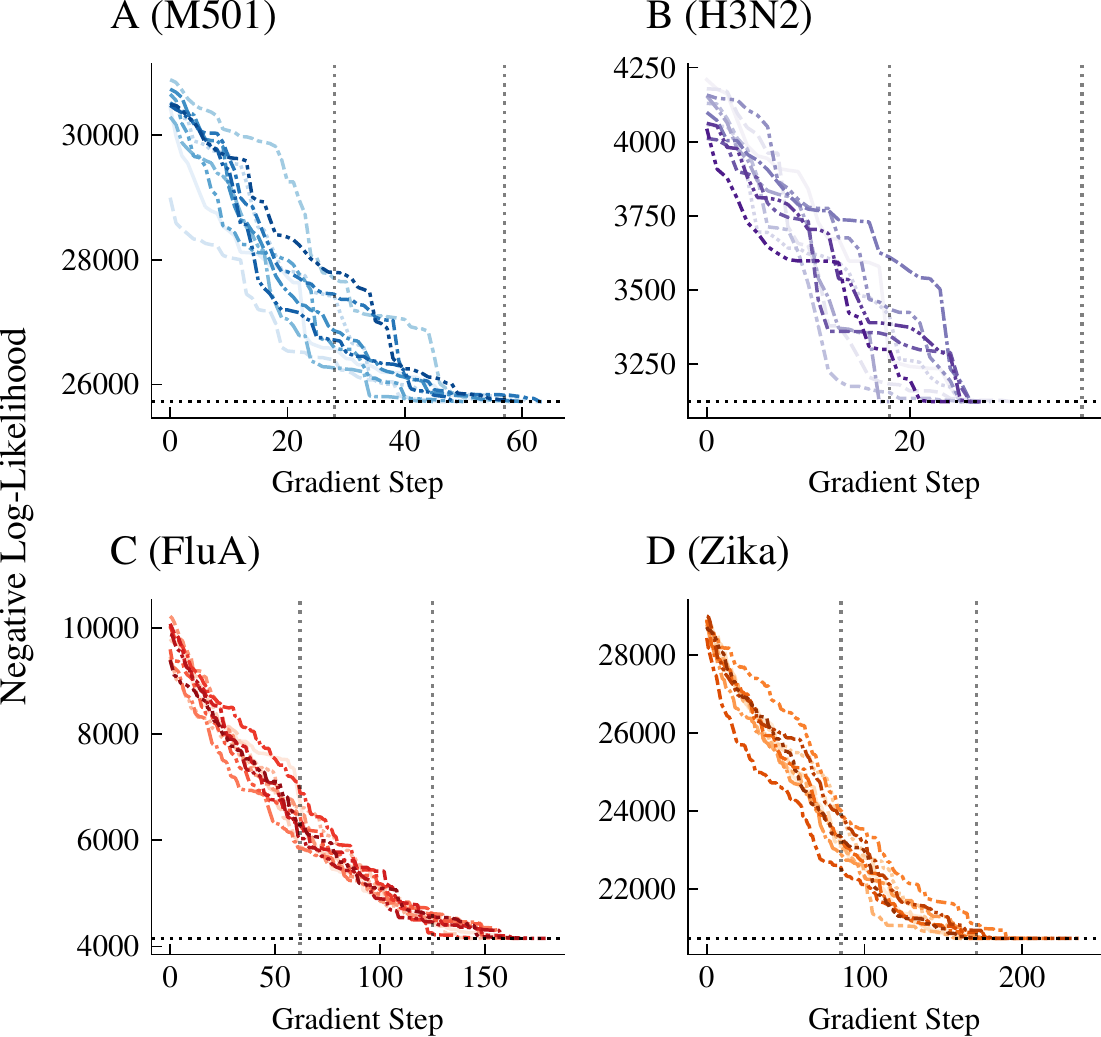}
    \caption{Phylo2Vec-based likelihood optimisation results for four datasets described in Table~\ref{tab:data}. The horizontal and vertical lines indicate local minima and epochs (i.e., one pass through every index of \v), respectively.}
    \label{fig:results}
    \vspace{2em}
    \includegraphics[width=0.75\linewidth]{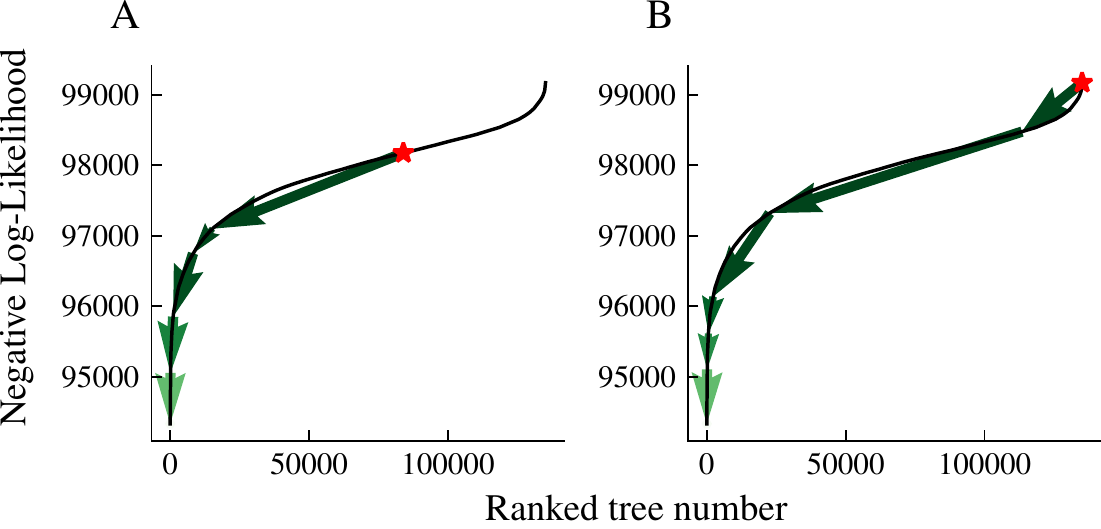}
    \caption{Negative log-likelihood path drawn from all possible trees of the Yeast dataset. A and B respectively show the path to the minimum from a random tree and the worst possible tree. The solid line shows the sorted phylogenetic likelihoods for all trees. The arrows show the proposal moves for two searches, one from a random tree (A) and one from the worst possible tree (B).}
    \label{fig:biggradient}
\end{figure}

Figure~\ref{fig:results} shows the optimisation results for four of the datasets described in Table~\ref{tab:data}. We observe that from 10 random starting trees, we always achieve the same minimal loss without being trapped in local optima. This is comparable to state-of-the-art software that also searches through topological space~\citep{stamatakis2014,Minh2020-ey}. For each dataset, only two epochs of changes (i.e., two passes through every index of \v) were generally needed to achieve a minimal negative log-likelihood. In addition, for M501 for example, only a total of around 10,000 likelihood optimisations for each run were needed to reach a minimum - a vanishingly small fraction of the total number of trees possible with 29 taxa ($\sim 8e^{36}$). The choice of the number of optimisations can be shortened depending on the optimisation stoppage criteria, but with the trade-off of being trapped in local minima. We also note that in the Zika virus example, two runs converged at a loss slightly (0.07\%) greater than the minimum of the other eight runs. The resultant trees from these minima show that we get trapped in these suboptimal minima due to rooting issues, preventing single changes in \v~from finding a better optimum. This highlights once again that our algorithm is attempting to solve a more difficult problem than is strictly necessary by searching the space of rooted trees rather than unrooted trees. Due to the pulley principle~\citep{felsenstein2004}, all rootings of an unrooted tree have the same negative log-likelihood and therefore no paths between rooted trees exist to aid our optimisation algorithm. In practice, especially for large phylogenies, it is common to begin optimisation from a sensible starting point~\citep{paradis2004} (e.g., a maximum parsimony or neighbour joining tree). In our experiments, we have chosen to start from a completely random tree to highlight the utility of simple algorithms based on Phylo2Vec to traverse tree space. 

Subsequently, we apply the same optimisation procedure for the yeast dataset (8 taxa) initially presented in~\citep{rokas2003} and studied in \citep{Money2012-ti}. Given the smaller number of taxa, we were able to exhaustively calculate the likelihood for every possible rooted tree. As shown in Figure~\ref{fig:biggradient}, we notice a broad region of numerous trees with comparable likelihoods, in addition to a considerably smaller group of trees exhibiting increasing likelihood. Regardless of whether we start from a random tree or the worst possible tree, our algorithm quickly converges to the accurate tree reported in~\citep{rokas2003}. Across several runs, Algorithm~\ref{alg:hill_climbing} required 96 total likelihood evaluations - a very small fraction of the total number of trees.

\section{Discussion}
Phylo2Vec is a parsimonious representation for phylogenetic trees whose validity extends to any binary tree. This representation facilitates the calculation of distances between trees and allows the construction of a simple algorithm for phylogenetic optimisation. Following from trends in phylogenetics, Phylo2Vec could be integrated within state-of-the-art computing libraries (e.g., \texttt{libpll}~\citep{flouri2015} or \texttt{Beagle}~\citep{ayres2012}) to facilitate its use. We have not yet considered Bayesian inference, but this is likely a useful application of Phylo2Vec, where random walks can be trivially implemented (see Figure~\ref{fig:distance}). Furthermore, Phylo2Vec can be useful in assessing topological convergence, for example, for a large phylogeny of 500 taxa and a million trees, extracting the unique set of topologies takes $<10$ seconds on a single core in Python, and can be even faster with parallel computation. Although Phylo2Vec does allow for unrooted trees, it is primarily an algorithm for rooted trees. In the examples in this paper, we only consider reversible Markov models where rooting is irrelevant due to the pulley principle~\citep{felsenstein2004}. Irreversible Markov models are both mathematically and biologically more principled \citep{Sumner2012-ps} but require rooted trees. Therefore, a useful application of Phylo2Vec could be in the inference of phylogenies with irreversible Markov models.

The use of empirical datasets served as a proof of concept that maximum likelihood estimation can be performed using Phylo2Vec vectors. We show that, using a simple hill-climbing scheme, we can recover the same topology optimum found by state-of-the-art MLE frameworks such as RAxML-NG~\citep{kozlov2019}. It is important to note, however, that this approach is nowhere near as optimised as RAxML-NG. As it only performs topology changes at a single vector index at a time, its inherent greediness makes inference of large datasets difficult.

That being said, the simplicity of the Phylo2Vec formulation means that it can be used in other more efficient and complex optimisation schemes can also be developed. For instance, Phylo2Vec can also benefit from fast SPR changes~\citep{guindon2010} and other heuristic optimisations that are currently in RAxML(-NG). In addition, by construction, we have ensured that Phylo2Vec can be differentiable through transforming $v$ into a matrix $W\in \mathbb{R}^{0,1}$ such that $W_{ij} = \mathbb{P}(\text{\v}_i = j)$. Via this transform, inference in a continuous tree space using gradient descent-based optimisation frameworks is theoretically possible, but its particulars remain to be developed. Similarly, we expect Phlyo2Vec-based representations to be applied in Monte Carlo tree search (MCTS) frameworks which may explore tree space more efficiently, or used as an embedding to regularly infer phylogenetic trees using well-established machine learning paradigms such as self-supervised learning from large existing tree libraries (e.g., TreeBase~\citep{piel2000}).

\section{Data and Code availability}
All code relevant to reproduce the experiments is available online: \url{https://github.com/Neclow/phylo2vec}. Instructions to access the publicly available datasets used in this study are included in the \texttt{phylo2vec/datasets} folder of the repository.

\section{Author contributions}
S.B, N.S, and M.J.P conceived the study. S.B supervised. S.B, N.S, and M.J.P designed the study. S.B, M.J.P, and N.S performed optimisation runs. S.B, M.J.P, and N.S performed analysis. S.B, M.J.P, and N.S drafted the first original draft. All authors contributed to editing the original draft. N.S and D.A.D contributed to revisions of the methodology. M.J.P, N.S, and S.B. drafted the appendix. N.S wrote the official implementation of Phylo2Vec.

\section{Competing interests}
The authors declare no competing interests.

\section{Funding}
M.J.P acknowledges support from his EPSRC DTP studentship, awarded by the University of Oxford to fund his DPhil in Statistics. D.A.D acknowledges support from the Novo Nordisk Foundation via The Data Science Investigator Award (NNF23OC0084647). S.B. acknowledges funding from the MRC Centre for Global Infectious Disease Analysis (reference MR/X020258/1), funded by the UK Medical Research Council (MRC). This UK funded award is carried out in the frame of the Global Health EDCTP3 Joint Undertaking.  S.B. is funded by the National Institute for Health and Care Research (NIHR) Health Protection Research Unit in Modelling and Health Economics, a partnership between UK Health Security Agency, Imperial College London and LSHTM (grant code NIHR200908). Disclaimer: “The views expressed are those of the author(s) and not necessarily those of the NIHR, UK Health Security Agency or the Department of Health and Social Care.” S.B. acknowledges support from the Novo Nordisk Foundation via The Novo Nordisk Young Investigator Award (NNF20OC0059309). S.B. acknowledges support from the Danish National Research Foundation via a chair grant (DNRF160) which also supports N.S. S.B. acknowledges support from The Eric and Wendy Schmidt Fund For Strategic Innovation via the Schmidt Polymath Award (G-22-63345). S.B and N.S acknowledge the Pioneer Centre for AI, DNRF grant number P1 as affiliate researchers. C.A.D receives support from the NIHR HPRU in Emerging and Zoonotic Infections, a partnership between the UK Health Security Agency, University of Liverpool, University of Oxford and Liverpool School of Tropical Medicine (grant code NIHR200907).



\section{Acknowledgements}
The authors thank Anthony Jakob (McKinsey \& Company) for proofreading the manuscript.

\clearpage



\clearpage
\normalsize

\section{Figure captions}
\markboth{Figure captions}{Figure captions}
\begin{enumerate}
    \item \nameref{fig:ordered}
    \item \nameref{fig:to_newick}
    \item \nameref{fig:examples_n4}
    \item \nameref{fig:to_vector}
    \item \nameref{fig:distance}
    \item \nameref{fig:reorder_v2}
    \item \nameref{fig:benchmarks}
    \item \nameref{fig:results}
    \item \nameref{fig:biggradient}
\end{enumerate}

\section{Table captions}
\markboth{Table captions}{Table captions}
\begin{enumerate}
    \item \nameref{tab:data}
\end{enumerate}




\clearpage
\setcounter{figure}{0}
\setcounter{table}{0}
\setcounter{algorithm}{0}
\makeatletter
\renewcommand{\thefigure}{S\@arabic\c@figure}
\renewcommand{\thetable}{S\@arabic\c@table}
\renewcommand{\thealgorithm}{S\@arabic\c@algorithm}
\makeatother

\markboth{Appendix}{Appendix}
\appendix
In this appendix, we present a rigorous mathematical framework summarising the results in this paper. We also detail further algorithms that may be useful in the implementation of Phylo2Vec.

\subsection{Notations and definitions}\label{sec:background}
\subsubsection*{Notations}
We shall use the notation $\mathcal{T}$ to refer to a tree, $\boldsymbol{b}$ for the branch lengths of $\mathcal{T}$ and $n$ as the number of leaves (or taxa). 

\subsubsection*{Node labels} In a tree with $n$ leaves, we set the convention that the leaf nodes are \emph{labelled} from $0$ to $n-1$, the internal nodes are \emph{labelled} from $n$ to $2n-3$ and the root (if it exists) is labelled as $2n-2$.

\subsubsection*{Generation}
By definition, there is a unique path connecting each pair of nodes in a tree. Suppose that the path from node $i$ to the root contains the nodes $i,a_1,\ldots,a_{g_i}$ in that order (and the root is hence $a_{g_i}$). We call $g_i$ the \textit{generation} of node $i$. The path from any $a_x$ to the root must be $a_x,a_{x+1},\ldots,a_{g_i}$ and hence $g_{a_x} = g_i - x$.

\subsubsection*{Unrooting}
Suppose that, in a rooted tree, the two children of the root are $x$ and $y$. To ``unroot” the tree, we remove the edges connecting $x$ and $y$ to the root and add a new edge connecting $x$ and $y$. This edge is given length $b_x + b_y$ where, for example, $b_x$ is the length of the original edge joining $x$ to the root. We shall use $\mathcal{T}'$ to refer to the unrooted tree and $\boldsymbol{b}'$ to its branch lengths.

\subsection{Phylo2Vec details}\label{sec:phylo2vec}
\subsubsection*{Vector representation}
As described in the main text, Phylo2Vec is a way to represent binary trees with $n$ leaves using a single integer vector, \v~of dimension $n - 1$ that is simply constrained by
\begin{equation*}
    \boldsymbol{v}[j] \in \{0, 1, \ldots, 2(j-1)\} ~\forall j \in \{1, \ldots,n-1\}
\end{equation*}
The construction of the tree from this representation and the bijectivity between \v~and the space of trees are covered in the main text.

\subsubsection*{Bijectivity of \v~to the space of all possible trees}
We can show that our mapping from $\mathbb{V}$ to the space of trees is a bijection.
\begin{lemma}
    \textit{The mapping between the set of vectors $\boldsymbol{v}\in \mathbb{V}$ and the set of (topologically equivalent, labelled) trees is a bijection.}
    \label{lemma:bijection}
\end{lemma}

\textbf{Proof:} As, by construction, the number of possible vectors \v~and the number of trees are the same ($(2n-3)!!$), it is simply necessary to show that this mapping is injective.

For any two nodes $a$ and $b$, we define $M(a,b)$ to be their most recent common ancestor (MRCA). Moreover, for any nodes $a$ and $b$, we use the notation $a \prec b$ if $a$ is an ancestor of $b$.

We can use the fact that the MRCA of any pair of leaf nodes in the tree is unchanged during the Phylo2Vec construction process (once both nodes have been added to the tree). Note that the path from a node $x$ to the root changes through the addition of an extra node if and only if a new node is appended to an edge on this path. Thus, if $M(a,b) = x$, $x$ will remain on the paths from $a$ and $b$ to the root. Moreover, nodes of higher generation than $x$ can only be added to one of the paths (as otherwise there would have to be an edge connecting at least one node of higher generation than $x$ on both the original paths, contradicting $x$ being the MRCA). Thus, the MRCA of $a$ and $b$ will be unchanged. Similarly, we can see that $a \prec b$ for two $a$ and $b$ at a given stage of the algorithm, this relationship will be unchanged throughout the construction process.

Suppose that two vectors \v~and $\boldsymbol{v}'$ result in trees $\mathcal{T}$ and $\mathcal{T}'$, and that $\boldsymbol{v} \neq \boldsymbol{v}'$. Define $i := \min\{j : v_j \neq v'_j\}$.

Define $X$ to be the set of leaf nodes that, just before node $i$ is added, are descended from the edge to which node $i$ is appended when constructing the tree according to $\boldsymbol{v}$, and $X'$ to be the equivalent set for $\boldsymbol{v}'$. We have $X \neq X'$, as each edge has a unique set of nodes descended from it. 

If both $X\setminus X'$ and $X'\setminus X$ are non-empty, suppose that $a \in X\setminus X'$ and $b \in X'\setminus X$. Then, in $\mathcal{T}$, it must be the case that $M(b,i) \prec M(a,i)$ (both at this stage in the algorithm, and in the final tree, by the previous argument) once node $i$ has been added (as the $M(a,i)$ will be the new internal node and $M(i,b)$ cannot be either $i$ or this new internal node). A similar argument shows that $M(a,i) \prec M(b,i)$ in $\mathcal{T}'$ and hence $\mathcal{T} \neq \mathcal{T}'$.

If only one of $X\setminus X'$ and $X'\setminus X$ is non-empty, suppose without loss of generality that $X\setminus X'$ is non-empty and choose $a \in X\setminus X'$. We can choose a distinct node $b \in X'\cap X$ (as otherwise, if $X' \cap X$ and $X' \setminus X$ are both empty, we must have one of $X$ and $X'$ being empty which is a contradiction as every edge has at least one leaf node descended from it). Then, $M(i,a)$ is the newly-added internal node in the construction of $\mathcal{T}$, but not in the construction of $\mathcal{T}'$. However, in both cases, $M(i,b)$ is the newly-added node. Hence, in $\mathcal{T}$, $M(i,a) = M(i,b)$, but in $\mathcal{T}'$, they are distinct. Thus, $\mathcal{T}' \neq \mathcal{T}$.

Hence, topological non-equivalence holds in both cases, and the map from \v~to the set of trees is therefore injective and therefore bijective.

\vspace{-0.5cm}
\subsubsection*{Label-asymmetry of $\boldsymbol{v}$-induced distance}
As discussed in the main text, $\boldsymbol{v}$ induces a natural distance function between trees - namely, that the distance between $\boldsymbol{v}$ and $\boldsymbol{w}$ is equal to $\sum_{i=0}^{k}\mathds{I}_{v_i \neq w_i}$.

However, this distance function is dependent on the labels assigned to each leaf (and is hence label-asymmetric). A simple example of this can be found in the case of four leaves.

Consider the tree given by $\boldsymbol{v}_1 = (0,1,2)$. This is a ladder tree (that is, each internal node and the root is parent to at least one leaf node) with nodes in order $0,1,\{2,3\}$ (where the $\{2,3\}$ is used to denote the fact that nodes 2 and 3 are from the same generation and so could be read in either order). This is distance 1 away from $\boldsymbol{v}_2 = (0,1,4)$, which is again a ladder tree with nodes now ordered as $3,0,\{1,2\}$. Thus, if distance were symmetric, then any ladder tree with ordered nodes $a,b,\{c,d\}$ would be distance 1 away from the ladder trees with ordered nodes $d,a,\{b,c\}$ and $c,a,\{b,d\}$. However, the ladder tree with ordered nodes $2,0,\{1,3\}$ is given by $\boldsymbol{v}_3 = (0,2,1)$, which is distance 2 away from $\boldsymbol{v}_1$. Thus, the distance function is not label-symmetric.

\subsubsection*{Unrooted tree equivalence classes}
Noting from the main text that there are $(2k-3)!!= \prod_{i=0}^{k-2}(2i+1)$ unrooted trees with $k+1$ leaves, we can partition the space of trees with $k+1$ leaves into $(2k-3)!!$ equivalence classes $\{\mathcal{E}_i : i = 1,2,\ldots,(2k-3)!!\}$ such that the removing the root from each of the trees in a given class $\mathcal{E}_i$ (according to the procedure described in \nameref{sec:background} in the Appendix) results in the same unrooted tree. These equivalence classes all contain $(2k-1)$ trees (as, reversing the unrooting procedure, a root can be added onto each of the $(2k-1)$ edges of an unrooted tree).

From the previous work, Felsenstein's likelihood is constant in each equivalence class. This has two consequences when attempting to maximise the likelihood. Firstly, a global minimum in the space of rooted trees will not be unique, as any of the other trees in the same equivalence class will also give a maximal likelihood. Secondly, if \v~is a local maximum (that is, all $\boldsymbol{w}$ with one entry different to \v~give a lower likelihood value), equivalence classes provide a way to change to a new vector $\boldsymbol{u}$ without having to decrease the likelihood.

Similarly to the case of reordering, the set of equivalence classes of vectors that are distance 1 from \v~will not in general be the same as the set with distance 1 from $\boldsymbol{u}$. One can either change equivalence class randomly or (unlike in the case of reordering), as there are only $2k-1$ members of each class, systematically iterate through them. Finding the vectors $\boldsymbol{w}$ corresponding to the trees in an equivalence class can again be done by constructing the tree matrix $M$, making the appropriate changes and then using the inversion algorithm to create a new vector representation.

Our experiments have shown that changing \v~is effective in increasing algorithmic performance. However, we found that reordering provided better increase in performance, and therefore did not present this other switching method in the algorithm in the main text.

\subsubsection*{Number of moves}
Let a vector \v~of length $n-1$ representing a tree with $n$ leaves.
For each entry $j$, there are $2j - 1 - 1 = 2 (j - 1)$ possible moves, as there are $2j-1$ possible entries for entry j, but \v$[j]$ is already set). Summing for all $j \in {1, \ldots, n-1}$ gives:
\begin{equation*}
    \sum_{j = 1}^{n-1} 2(j-1) = (n-1)(n-2) \approx n^2
\end{equation*}

Thus, the number of possible moves for Phylo2Vec is of the order $\mathcal{O}(n^2)$.

\clearpage
\subsection{Algorithms}\label{sec:algorithms}
The algorithms provided here are presented in rough descriptive pseudocode; they are written in human-readable fashion to facilitate comprehension, but are not necessarily optimised. Please refer to the GitHub repository (\url{https://github.com/Neclow/phylo2vec}) for detailed implementation.

\begin{algorithm}[h!]
\caption{Labelling a rooted tree as an ordered \v}
\label{alg:alg0}
\begin{algorithmic}
\State \textbf{Input} rooted tree $\mathcal{T}$ with $n$ leaves
\State $\text{\v} \gets [0]$ \Comment{Initial integer vector}
\ForAll{leaves $j = 2, ..., n-1$}
    \State $\mathcal{T}_j \gets$ subtree with leaves $0, ..., j$
    \State $s \gets $ leaf in $\mathcal{T}_j$ such that $j$ and $s$ form a cherry
    \State \v.append($s$)
\EndFor
\State \Return \v
\end{algorithmic}
\label{alg:ordered_to_vector}
\end{algorithm}

\begin{algorithm}[h!]
\caption{Recovering an ordered rooted tree from an ordered \v}
\label{alg:alg0_inverse}
\begin{algorithmic}
\State \textbf{Input} \v~of size $n-1$ satisfying Eq.~\ref{eq:ordered} \Comment{Ordered integer vector}
\State $\mathcal{T} \gets$ a rooted tree with two leaves 0 and 1 \Comment{Initial tree}
\ForAll{$j = 2, ..., n-1$}
    \State Add a new leaf $j$ to $\mathcal{T}$ such that it forms a cherry with \v$[j]$
    \State $a$($j$, \v$[j]$) $\gets 2(n-1) - j + 1$ \Comment{Ancestor of $j$ and \v$[j]$}
\EndFor
\State \Return $\mathcal{T}$
\end{algorithmic}
\label{alg:ordered_to_newick}
\end{algorithm}

\begin{algorithm}[htbp]
\caption{Recovering a rooted tree from a Phylo2Vec \v}
\begin{algorithmic}
\State \textbf{Input} \v~of dimension $n-1$ satisfying Eq.~\ref{eq:phylo2vec} \Comment{Phylo2Vec vector}
\State pairs $\gets [(0, 1)]$ \Comment{\parbox[t]{.5\linewidth}{The objective of the algorithm is find the nodes to merge and their order. The first pair is always (0,1) for a 2-leaf tree.}}

\ForAll{leaves $j = 2, ..., n-1$}
    \If{\v$[j] \leq j$} \Comment{\parbox[t]{.5\linewidth}{This is like an ordered vector. The branch leading to \v[j] gives birth birth to leaf $j$}}
        \State next pair $\gets (\text{\v}[j], j)$
        \State position $\gets$ 1 \Comment{\parbox[t]{.5\linewidth}{This pair corresponds to a cherry at height 0, so it should be drawn first.}}
    ~\\
    \Else \Comment{The branch to be split is internal}
        \State position $\gets \text{\v}[j]$ - len(pairs) + 1 \Comment{\parbox[t]{.5\linewidth}{The index at which the next pair will be inserted}}
        \State descendant $\gets$ pairs[position - 1][1] \Comment{\parbox[t]{.5\linewidth}{A node that we processed beforehand that is deeper than the branch \v$[j]$}}
        \State next pair $\gets$ (pairs[descendant, $j$]) 
    \EndIf
    \State pairs.insert(position, next pair)
\EndFor

\State \Return pairs \Comment{\parbox[t]{.5\linewidth}{The pairs indicate how to build the tree (and form a Newick string)}}



\end{algorithmic}
\label{alg:to_newick}
\end{algorithm}

\begin{algorithm}[t]
\caption{Labelling a rooted tree as a Phylo2Vec vector}
\begin{algorithmic}
\State \textbf{Input} rooted tree $r$ with $n$ leaves \Comment{\parbox[t]{.6\linewidth}{We assume a Newick string with integer nodes and labelled internal nodes. \\ Ex: (((2,3)6,1)7,(0,4)5)8;}}
\State $M \gets$ reduce($r$) \Comment{\parbox[t]{.6\linewidth}{``Reduce" the Newick string to a $(n-1)\times3$ matrix. \\ Columns 1-2 = children nodes \\ Column 3 = ancestor \\ Ex: $\begin{matrix}
    0 & 4 & 5 \\
    2 & 3 & 6 \\
    1 & 6 & 7 \\
    5 & 7 & 8 \\
\end{matrix}$
}} 
\State $C \gets$ extract\_cherries($M$) \Comment{\parbox[t]{.6\linewidth}{Starting from the smallest internal node, replace the internal nodes by their smallest child (and discard the third column). This is equivalent to the pairs in Algorithm~\ref{alg:to_newick} \\ Ex: $\begin{matrix}
    0 & 4 \\
    2 & 3 \\
    1 & 2 \\
    0 & 1 \\
\end{matrix}$}} 
\State $\text{\v} \gets$ build\_vector($C$) \Comment{\parbox[t]{.6\linewidth}{Use the same logic as Algorithm~\ref{alg:to_newick} to retrieve each \v$_j$ from the position of the containing leaf $j$}}

\State \Return \v
\end{algorithmic}
\label{alg:to_vector}
\end{algorithm}




\clearpage
\subsection{Supplementary Figures}\label{sec:supp_figures}

\begin{figure}[htbp]
    \centering
    \includegraphics[width=0.6\linewidth]{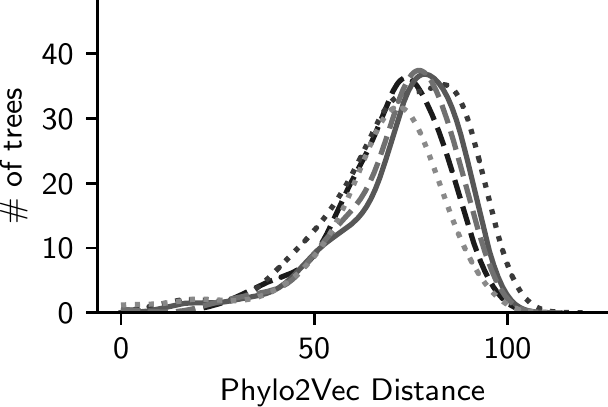}
    \caption{Density plots of Phylo2Vec distance $\mu$ for an equivalence set of rooted trees with 200 taxa with a fixed Phylo2Vec vector \v. 5 random \v~are shown.}
    \label{fig:equiv_class}
\end{figure}

\begin{figure}[htbp]
    \centering
    \subfloat[]{\includegraphics[width=0.48\linewidth]{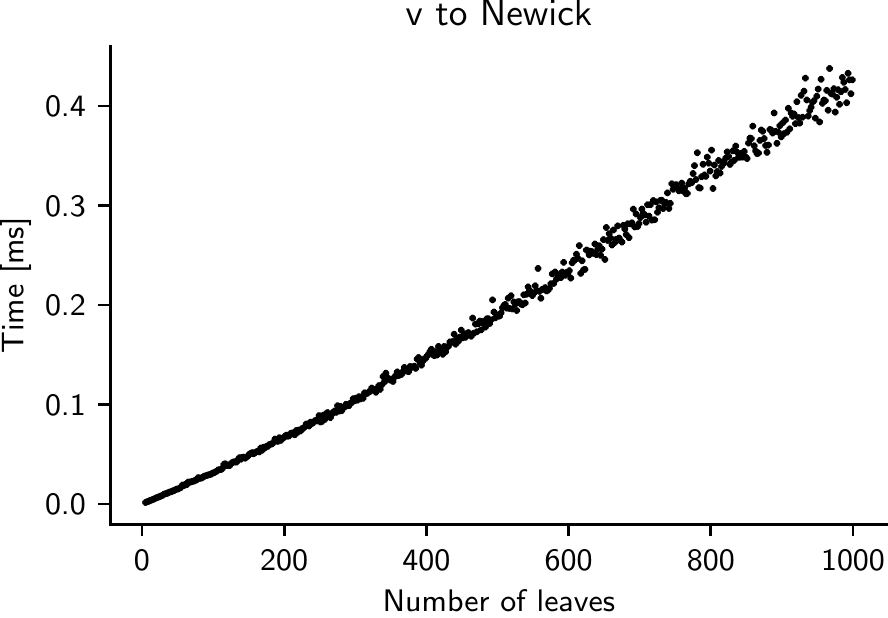} \label{fig:cpxity_to_newick}}
    \subfloat[]{\includegraphics[width=0.48\linewidth]{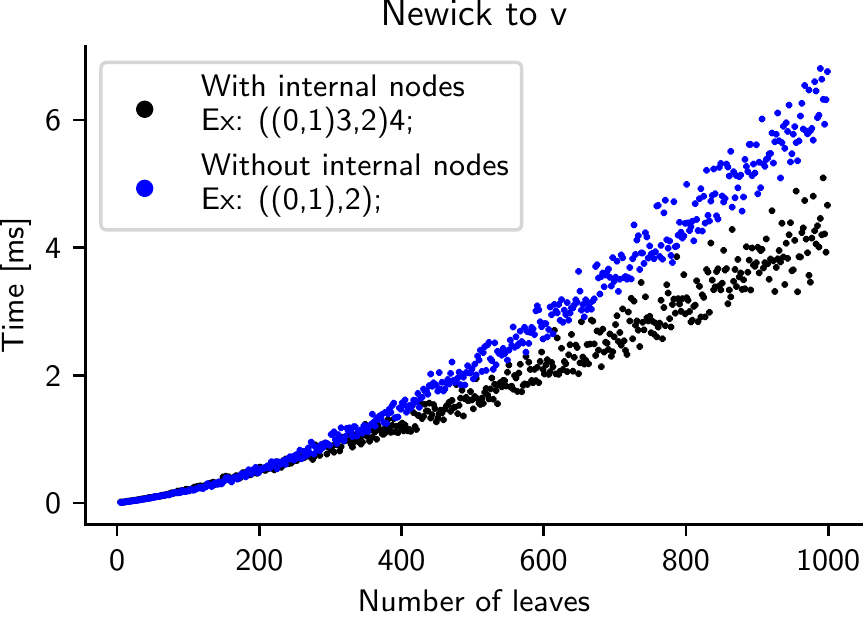} \label{fig:cpxity_to_vector}}
    \caption{Execution times for different tree sizes of \textsb{(a)} Algorithm~\ref{alg:to_newick} to recover a rooted tree from a Phylo2Vec \v~and \textsb{(b)} Algorithm~\ref{alg:to_vector} to label a rooted tree as a Phylo2Vec \v. For each size, we evaluated the execution time with a fixed configuration of 100 executions in a loop and 7 repeats using \texttt{timeit} in Python. When the number of taxa is large, manipulations on long strings to produce or process the Newick string can increase memory and thus increase execution time.}
 \label{fig:cpxity}
\end{figure}

\end{document}